\documentclass[acmsmall]{acmart}

\AtBeginDocument{%
  \providecommand\BibTeX{{%
    \normalfont B\kern-0.5em{\scshape i\kern-0.25em b}\kern-0.8em\TeX}}}

\usepackage{graphicx, url}
\usepackage{balance}
\usepackage{subfigure,captcont,wrapfig, setspace}
\usepackage{comment, ragged2e}
\usepackage{subfigure, color, colortbl}
\usepackage{amsmath}
\usepackage[linesnumbered, ruled, vlined]{algorithm2e}

\setcopyright{acmcopyright}
\copyrightyear{2020}
\acmYear{2020}
\acmDOI{0000001.0000001}

\acmJournal{TECS}
\acmVolume{0}
\acmNumber{0}
\acmArticle{0}
\acmMonth{8}

\definecolor{blue}{rgb}{0.0,0.0,1}

\definecolor{red}{rgb}{1,0.0,0.0}

\newcommand{\code}[1]{\texttt{#1}}

\begin{document}

\title{LPWAN in the TV White Spaces: A Practical Implementation and Deployment Experiences}

\author{Mahbubur Rahman}
\orcid{0000-0003-2353-6687}
\affiliation{%
  \institution{City University of New York}
  \streetaddress{Queens College Science Building, 65-30 Kissena Boulevard}
  \city{Flushing}
  \state{NY}
  \postcode{11367}
  \country{USA}}
\email{mdmahbubur.rahman@qc.cuny.edu}

\author{Dali Ismail}
\affiliation{%
 \institution{Wayne State University}
  \streetaddress{5057 Woodward Avenue, Suite 3010}
  \city{Detroit}
  \state{MI}
  \postcode{48202}
  \country{USA}}
\email{dali.ismail@wayne.edu}

\author{Venkata P. Modekurthy}
\affiliation{%
 \institution{Wayne State University}
  \streetaddress{5057 Woodward Avenue, Suite 3010}
  \city{Detroit}
  \state{MI}
  \postcode{48202}
  \country{USA}}
\email{modekurthy@wayne.edu}

\author{Abusayeed Saifullah}
\affiliation{%
 \institution{Wayne State University}
  \streetaddress{5057 Woodward Avenue, Suite 3010}
  \city{Detroit}
  \state{MI}
  \postcode{48202}
  \country{USA}}
\email{saifullah@wayne.edu}

\renewcommand{\shortauthors}{Trovato and Tobin, et al.}

\begin{abstract}
Low-Power Wide-Area Network (LPWAN) is an enabling Internet-of-Things (IoT) technology that supports long-range, low-power, and low-cost connectivity to numerous devices. To avoid the crowd in the limited ISM band (where most LPWANs operate) and cost of licensed band, the recently proposed SNOW (Sensor Network over White Spaces) is a promising LPWAN platform that operates over the TV white spaces. As it is a very recent technology and is still in its infancy, the current SNOW implementation uses the USRP devices as LPWAN nodes, which has high costs ($\approx$ \$750 USD per device) and large form-factors, hindering its applicability in practical deployment. In this paper, we implement SNOW using low-cost, low form-factor, low-power, and widely available commercial off-the-shelf (COTS) devices to enable its practical and large-scale deployment. Our choice of the COTS device (TI CC13x0: CC1310 or CC1350) consequently brings down the cost and form-factor of a SNOW node by 25x and 10x, respectively. Such implementation of SNOW on the CC13x0 devices, however, faces a number of challenges to enable link reliability and communication range. Our implementation addresses these challenges by handling peak-to-average power ratio problem, channel state information estimation, carrier frequency offset estimation, and near-far power problem. Our deployment in the city of Detroit, Michigan demonstrates that CC13x0-based SNOW can achieve uplink and downlink throughputs of 11.2kbps and 4.8kbps per node, respectively, over a distance of 1km. Also, the overall throughput in the uplink increases linearly with the increase in the number of SNOW nodes.
\end{abstract}

\ccsdesc[500]{Networks~Network architectures}
\ccsdesc[500]{Networks~Network performance evaluation}
\ccsdesc[300]{Networks~Network mobility}
\ccsdesc[100]{Networks~Network protocol design}
\ccsdesc[300]{Computer systems organization~Sensor networks}

\keywords{LPWAN, SNOW, White spaces, OFDM}

\maketitle
\renewcommand{\shortauthors}{Rahman, M. et al.}

\section{Introduction}\label{sec:intro}

Low-Power Wide-Area Network (LPWAN) is an emerging communication technology that supports long-range, low-power, and low-cost connectivity to numerous devices. It is regarded as a key technology to drive the Internet-of-Things (IoT). Due to their escalating demand, recently multiple LPWAN technologies have been developed that operate in the licensed/cellular (NB-IoT, LTE-M, 5G, etc.) or unlicensed/non-cellular (SNOW, LoRa, SigFox, etc.) bands~\cite{ismail2018low}. Most of the non-cellular technologies operate in the sub-1GHz ISM band except SNOW (Sensor Network over White Spaces) and WEIGHTLESS-W that operate in the TV white spaces~\cite{ismail2018low, whitespaceSurvey}.

\emph{White spaces} are the allocated but locally unused TV spectrum (54-698MHz in the US) that can be used by unlicensed devices as the secondary users.
Compared to the crowded ISM band, white spaces offer less crowded and much wider spectrum in both urban and rural areas, boasting an abundance in rural and suburbs~\cite{snow2}. Due to their low frequency, white spaces have excellent propagation and obstacle penetration characteristics enabling long-range communication. Thus, they hold the potentials for LPWAN to support various IoT applications. 
To our knowledge, WEIGHTLESS-W (which, to the best of our knowledge, has been decommissioned~\cite{ismail2018low}) and SNOW~\cite{snow_ton} are the only two efforts to exploit the TV white spaces for LPWAN. 
Initially introduced in~\cite{snow}, SNOW is a highly scalable LPWAN technology offering reliable, bi-directional, concurrent, and asynchronous communication between a base station (BS) and numerous nodes.


Despite its promise as a LPWAN technology, SNOW has not yet received sufficient attention from the research community due to its limited availability for practical deployment.
The SNOW implementation in~\cite{snow_ton}, which is also available as open-source~\cite{snow_bs}, uses Universal Software Radio Peripheral (USRP) devices as LPWAN nodes, hindering the applicability of this technology in practical and large-scale deployment.
USRP is a hardware platform developed for software-defined radio applications~\cite{usrp}. Using them as the SNOW node limits the practical deployment of SNOW in real-world applications due to several factors including its high cost and large form-factor. Currently, a USRP B200 device with a half-duplex radio costs $\approx$ \$750 USD. As such, it inherently becomes costly to deploy a large-scale SNOW network. 
Today, IoT applications including smart city (e.g., waste management and smart grid), transportation and logistics (e.g., connected vehicles), agricultural and smart farming (i.e., Microsoft FarmBeats), process management (e.g.,oil field monitoring), and healthcare require collection of information from thousands of IoT nodes~\cite{ismail2018low}.

In this paper, we address the above practical limitations of the existing SNOW technology by implementing it on low-cost and low form-factor commercial off-the-shelf (COTS) devices.
Through this implementation, we demonstrate that any COTS device that has a programmable physical layer (PHY), operates in the white spaces, and supports amplitude-shift-keying (ASK) or binary phase-shift-keying (BPSK) may work as a SNOW node in practical deployments. Along with our original TI CC1310-based SNOW implementation in~\cite{snow_cots}, we thus implement SNOW on TI CC1350 that has a programmable PHY, costs approximately \$30 USD (retail price) and is 10x smaller than a USRP B200 device, thereby making SNOW adoptable in practical IoT applications. Additionally, we emphasize the fact that this implementation maybe is adaptable in the diverse range of COTS IoT devices with minimal efforts (e.g., only needs to deal with new development software), which makes it highly portable and practical choice for SNOW.

The existing USRP-based SNOW implementation does not face the following practical challenges due to the expensive and powerful hardware design of USRP (as reflected by evaluation in~\cite{snow_ton, snow, snow2}), which the implementation on CC13x0 (CC1310 or CC1350) has to address. 
{\bf First}, due to its orthogonal frequency division multiplexing (OFDM)-based design, the SNOW BS transmitter is subject to high peak-to-average power ratio (PAPR). Thus, the overall reliability at the CC13x0 device during downlink communication degrades severely. 
{\bf Second}, due to the asymmetric bandwidth requirements of the SNOW BS and the nodes, channel state information (CSI) estimation between the BS and a CC13x0 device plays a vital role in both uplink and downlink communications. Without CSI estimation, the overall reliability and the communication range decreases. 
{\bf Third}, Carrier frequency offset (CFO) needs to be handled robustly as its effects are much more pronounced in the low-cost CC13x0 devices, leading to severe inter-carrier-interference (ICI). ICI decreases the overall bitrate in both uplink and downlink communications. 
While these challenges are quite common in the wireless domain, due to the novel design of SNOW, the existing solutions/approaches may not be adopted in SNOW.
Along with addressing these challenges, through this new implementation, we also make SNOW resilient to the classic near-far power problem. Due to the near-far power problem, where a far node's transmission gets buried under a near node's transmission radiation, the reliability in the uplink communication may be degraded. We thus address the above challenge as well.
Specifically, we make the following key contributions.
\begin{itemize}
	\item We implement SNOW for practical deployment on CC13x0s to work as SNOW nodes. Compared to the current USRP-based SNOW implementation, the cost and the form-factor of a single CC13x0-based SNOW node are decreased approximately 25x and 10x, respectively.
	\item In our implementation, we address several practical challenges including the PAPR problem, CSI and CFO estimation, and near-far power problem. Specifically,
	we propose a data-aided CSI estimation technique that allows a CC13x0 device to communicate directly with the SNOW BS from a distance of approximately 1km. Additionally, we propose a pilot-based CFO estimation technique that takes into account the device mobility and increases reliability in both uplink and downlink communications. Finally, we address the near-far power problem in SNOW through an adaptive transmission power control (ATPC) protocol that improves the reliability in the uplink communications.
	\item We experiment with the CC13x0-based SNOW implementation through deployment in the city of Detroit, Michigan. Our results demonstrate that we achieve an uplink throughput of 11.2kbps per node.
	Additionally, our overall uplink throughput increases \emph{linearly} with the increase in the number of SNOW nodes. In the downlink, we achieve a throughput of 4.8kbps per node.
	Compared to a typical LoRa deployment (channel bandwidth: 500kHz, spreading factor: 7, and coding rate: 4/5), our uplink throughput is approximately 3.7x higher when 5 nodes transmit to a gateway that can receive concurrent packets using 3 channels.
\end{itemize}



In the rest of the paper, Section~\ref{sec:model} provides an overview of SNOW and TI CC13x0. Section~\ref{sec:implementation} presents our SNOW implementation. Section~\ref{sec:near-far} describes the near-far power problem in SNOW. Sections~\ref{sec:deploy} and~\ref{sec:eval}  analyze the deployment cost and performance of our CC13x0-based SNOW, respectively. Section~\ref{sec:related} overviews related work. Finally, Section~\ref{sec:conclusion} concludes our paper.

\section{Background and System Model}\label{sec:model}
\subsection{An Overview of SNOW}\label{sec:snow_overview}
\begin{figure*}[!htbp]
    \centering
      \subfigure[Network Architecture\label{fig:network}]{
    \includegraphics[width=0.4\textwidth]{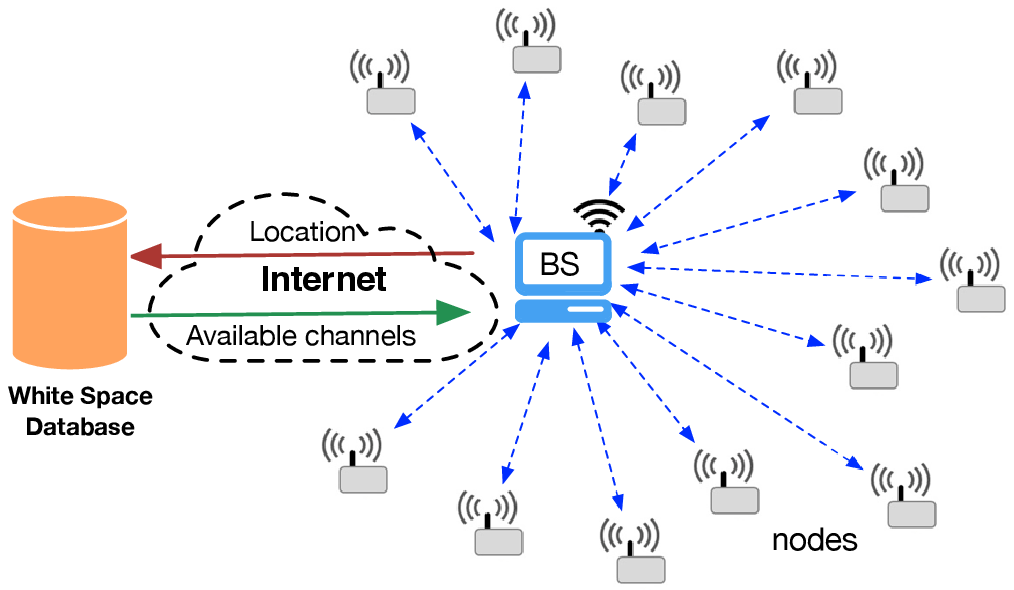}
      }\hfill
      \subfigure[Dual-radio BS and subcarriers\label{fig:dualradio}]{
        \includegraphics[width=.4\textwidth]{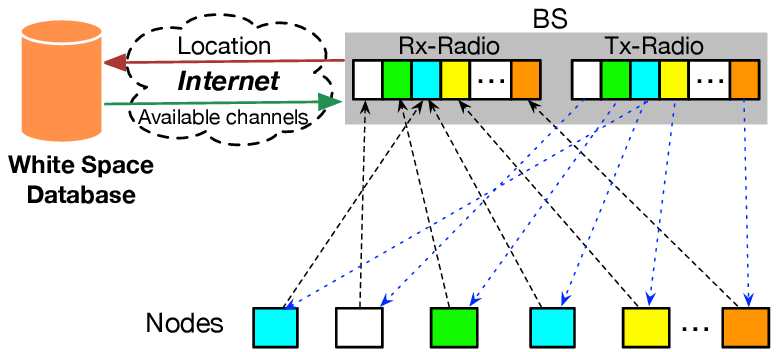}
      }
    \caption{The SNOW architecture~\cite{snow2}.}
    \label{fig:arch}
\end{figure*}
In this section, we provide a brief overview of SNOW. Its complete design and description is available in~\cite{mydissertation}. SNOW is a highly scalable LPWAN technology operating in the TV white spaces. It supports asynchronous, reliable, bi-directional, and concurrent communication between a BS and numerous nodes. Due to its long-range, SNOW forms a star topology allowing the BS and the nodes to communicate directly, as shown in Figure~\ref{fig:network}. The BS is powerful, Internet-connected, and line-powered while the nodes are power-constrained and do not have access to the Internet. To determine white space availability in a particular area, the BS queries a cloud-hosted geo-location database via the Internet. A node depends on the BS to learn its white space availability. In SNOW, all the complexities are offloaded to the BS to make the node design simple. 
Each node is equipped with a single half-duplex radio. To support simultaneous uplink and downlink communications, the BS uses a dual-radio architecture for reception (Rx) and transmission (Tx), as shown in Figure~\ref{fig:dualradio}.

The SNOW PHY uses a distributed implementation of OFDM called {\em D-OFDM}. D-OFDM enables the BS to receive concurrent transmissions from {\em asynchronous} nodes using a single-antenna radio (Rx-radio). Also, using a single-antenna radio (Tx-Radio), the BS can transmit different data to different nodes concurrently~\cite{snow, snow2, snow_ton, snow3, isnow_ton, isnow_p2p}. Note that the SNOW PHY is different from MIMO radio design adopted in other wireless domains such as LTE, WiMAX, and 802.11n~\cite{snow2} as the latter use multiple antennas to enable multiple transmissions and receptions.
The BS operates on a wideband channel split into orthogonal narrowband channels/subcarriers (Figure~\ref{fig:dualradio}). Each node is assigned a single subcarrier. 
For encoding and decoding, the BS runs inverse fast Fourier transform (IFFT) and global fast Fourier transform (G-FFT) over the entire wideband channel, respectively.
When the number of nodes is no greater than the number of subcarriers, every node is assigned a unique subcarrier. Otherwise, a subcarrier is shared by more than one node. 
SNOW supports ASK and BPSK modulation techniques, supporting different bitrates. 



The nodes in SNOW use a lightweight CSMA/CA (carrier sense multiple access with collision avoidance)-based media access control (MAC) protocol similar to TinyOS~\cite{tinyos}. The nodes can autonomously transmit, remain in receive mode, or sleep. Specifically, when a node has data to send, it wakes up by turning its radio on. Then it performs a random back-off in a fixed initial back-off window. When the back-off timer expires, it runs CCA (Clear Channel Assessment). If the subcarrier is clear, it transmits the data. If the subcarrier is occupied, then the node makes a random back-off in a fixed congestion back-off window. After this back-off expires, if the subcarrier is clean the node transmits immediately. This process is repeated until it makes the transmission and gets an acknowledgment (ACK).

\subsection{An Overview of TI CC13x0 LaunchPads}
Texas Instruments introduced TI CC1310 and TI CC1350 LaunchPads as a part of the SimpleLink microcontroller (MCU) platform to support ultra-low-power and long-range communication having a Cortex-M3 processor with 8KB SRAM and up to 128KB of in-system programmable storage~\cite{cc1350, snow_cots}.
With a small form-factor (length: 8cm, width: 4cm), both CC1310 and CC1350 are designed to operate in the lower frequency bands (287--351MHz, 359--527MHz, and 718--1054MHz) including the TV band. As an added feature, CC1350 can also operate in the 2.4GHz band while using Bluetooth low energy radio.
The packet structure of the CC13x0 devices includes a \code{preamble}, followed by \code{sync word}, \code{payload length}, \code{payload}, and \code{CRC}, chronologically. They support different data modulation techniques including Frequency Shift Keying (FSK), Gaussian FSK (GFSK), On-Off Keying (OOK), and a proprietary long-range modulation. They are capable of using a Tx/Rx bandwidth that ranges between 39 and 3767kHz. 
Additionally, with a supply voltage in the range of 1.8 to 3.8 volts, their Rx and Tx current consumption is 5.4mA and 13.4mA at +10dBm, respectively, offering ultra-low-power communication. 
The greatest advantage is that  
they have a programmable and reconfigurable physical layer, offering flexibility and feasibility for customized protocol implementation.

\section{SNOW Implementation on TI CC13x0}\label{sec:implementation}
\begin{figure}[!htb]
\centering
\includegraphics[width=0.49\textwidth]{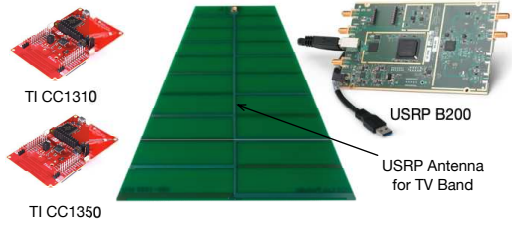}
\caption{Devices used in our SNOW implementation. A node is a CC1310 or CC1350 device. The BS has two USRP B200s, each having its own antenna. An antenna is approximately 2x bigger than a B200.}
\label{fig:devices}
\end{figure}
The original SNOW implementation in~\cite{snow_ton} uses the USRP hardware platform for both the BS and the nodes. In our implementation, we use the CC13x0 devices as SNOW nodes and USRP in the BS (Figure~\ref{fig:devices}).
For BS implementation, we adopt the open-source code provided in~\cite{snow_bs}. The BS uses two half-duplex USRP devices (Rx-Radio and Tx-Radio), each having its own antenna. Also. the BS is implemented on the GNURadio software platform that gives a high magnitude of freedom to perform baseband signal processing~\cite{gnuradio}.
In the following, we explore a number of implementation considerations and feasibility for a CC13x0 device to work as a SNOW node in practical deployments. 
First, we show how to configure a CC13x0 device to make it work as a SNOW node. We then address the practical challenges (e.g., PAPR problem, CSI estimation, and CFO estimation) associated with our CC13x0-based SNOW implementation.

\subsection{Configuring TI CC13x0}
We configure the subcarrier center frequency, bandwidth, modulation, and the Tx power by setting appropriate values to the CC13x0 command inputs \code{centerFreq, rxBw, modulation}, and \code{txPower}, respectively, using {\em Code Composer Studio} (CCS) provided by Texas Instruments~\cite{snow_cots}. A graphical user interface alternative to CCS is {\em SmartRF Studio}. The MAC protocol of SNOW in CC13x0 is implemented on top of the example CSMA/CA project that comes with CCS. Note that the functionalities of a SNOW node are very simple and may be incorporated easily in the IoT devices that have both storage and computational limitations like the CC13x0 devices.

\subsection{Peak-to-Average Power Ratio Observation}\label{sec:papr}
By transmitting on a large number of subcarriers simultaneously (in the downlink), the BS suffers from a traditional OFDM problem called {\em peak-to-average power ratio (PAPR)}. PAPR of an OFDM signal is defined as the ratio of the maximum instantaneous power to its average power.
In the SNOW downlink communications (i.e., BS to nodes), after the IFFT is performed by the BS, the composite signal can be represented as
$\nonumber x(t) = \frac{1}{\sqrt{N}}\sum_{k=0}^{N-1}X_k~e^{j2 \pi f_k t},~~0 \le t \le NT.$
Here, $X_k$ is the modulated data symbol for node $k = \{0, 1, \cdots, N-1\}$ on subcarrier center frequency $f_k = k\Delta f$, where $\Delta f = \frac{1}{NT}$ and $T$ is the symbol period. Therefore, the PAPR may be calculated as
\begin{equation}
\nonumber \text{PAPR}[x(t)] = 10\log_{10}\Bigg( \frac{\max\limits_{0~ \le ~t~ \le~ NT} [|x(t)|^2 ]}{P_{\text{avg}}}\Bigg)~~dB.
\end{equation}
Here, the average power $P_{\text{avg}} = E [|x(t)|^2]$.
A node's signal detection on its subcarrier is very sensitive to the nonlinear signal processing components used in the BS, i.e., the digital-to-analog converter (DAC) and high power amplifier (HPA), which may severely impair the bit error rate (BER) in the nodes due to the induced spectral regrowth. If the HPA does not operate in the linear region with a large power back-off due to high PAPR, the out-of-band power will exceed the specified limit and introduce severe ICI~\cite{jiang2008overview}. Moreover, the in-band distortion (constellation tilting and scattering) due to high PAPR may cause severe performance degradation~\cite{kamali2012understanding}. It has been shown that the PAPR reduction results in significant power saving at the transmitters~\cite{baxley2004power}.
\begin{figure}[!htb]
\centering
\includegraphics[width=0.35\textwidth]{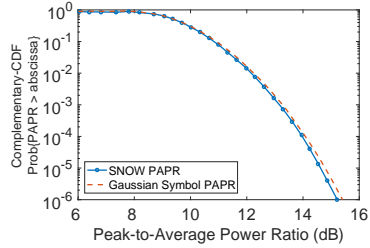}
\caption{PAPR distribution of D-OFDM signal in Tx-Radio.}
\label{fig:papr}
\end{figure}

As shown in Figure~\ref{fig:papr}, the PAPR in the SNOW downlink communications (for N = 64) follows the Gaussian distribution. Thus, the peak signal occurs quite rarely and the transmitted D-OFDM signal will cause the HPA to operate in the nonlinear region, resulting in a very inefficient amplification. To illustrate the power efficiency of the HPA for N = 64, let us assume the probability of the clipped D-OFDM frames is less than 0.01\%. We thus need to apply an input back-off (IBO)~\cite{baxley2004power} equivalent to the PAPR at a probability of $10^{-4}$. Here, PAPR $\approx$ 14dB or 25.12. Thus, the efficiency ($\eta = 0.5/\text{PAPR}$) of the HPA~\cite{jiang2008overview} is $\eta = 0.5/25.12 \approx 1.99\%$. Such low efficiency at the HPA motivates us to explore the high PAPR in SNOW for practical deployments.
Several uplink PAPR reduction techniques for single-user OFDM systems have been proposed (see survey~\cite{jiang2008overview}). However, the characteristics of the downlink PAPR in SNOW, where different data are concurrently transmitted to different nodes, are entirely different from the PAPR observed in a single-user OFDM system. To adopt an uplink PAPR reduction technique used in the single-user OFDM systems for the downlink PAPR reduction in SNOW, each node has to process the entire data frame transmitted by the BS and then demodulate its own data. However, a SNOW node has less computational power and does not apply FFT to decode its data~\cite{snow_ton}, or any other node's data. Thus, the existing PAPR reduction techniques will not work in our implementation.

We propose to handle the PAPR problem in SNOW by using only one subcarrier (called {\em downlink subcarrier}) for downlink communication. All the nodes use this subcarrier to receive from the BS. Namely, the Tx-Radio transmits only on one subcarrier that is not used by any node for uplink communication.
The BS may send any broadcast message, ACK, or data to the nodes using that downlink subcarrier. A node has to switch to the downlink subcarrier to listen to any broadcast message, ACK, or data.
The BS may reserve multiple subcarriers  as {\slshape backup subcarriers} for downlink communication. 
If the currently used downlink subcarrier becomes overly noisy or unreliable, it can be replaced by a backup subcarrier.
Note that the dual-radio in the BS allows it to receive concurrent packets from a set of nodes (uplink) and transmit broadcast/ACK/data packets to another set of nodes (downlink), simultaneously. 
The BS can acknowledge several nodes using a single transmission by using a bit-vector of size equals to the number of subcarriers.
If the BS receives a packet from a node operating on subcarrier $i$, it will set the $i$-th bit in the bit-vector. Upon receiving the bit-vector, that node may get an ACK by looking at the $i$-th bit of the vector. Because of the bit-vector, the downlink ACKs also scale up like the uplink traffic. In the case of different packets for different nodes, the volume of downlink traffic (compared to the uplink traffic) is also practical since the IoT applications may not require high volume downlink traffic~\cite{whitespaceSurvey}.



When a node $u$ transmits to the BS, if another node $v$ sharing the same subcarrier wants to transmit, $v$ senses the channel as busy and refrain from transmitting. When the BS transmits ACK to $u$ on the downlink subcarrier using the Tx-Radio, node $v$ may also transmit to the BS. Since the Tx-Radio at that time is making a downlink transmission, it may not send the ACK upon $v$'s transmission immediately. However, the Tx-Radio can send $v$'s ACK immediately after completing its current downlink transmission. Thus, $v$ may need to wait for ACK for a little longer than the time needed to send a downlink transmission from the BS. A node may go to sleep mode or its next state right after receiving an ACK. However, if a node that has transmitted but not yet received ACK, should wait for a little longer (e.g., up to one or two downlink transmission time). Note that a very few nodes (sharing the same subcarrier) may be involved in this scenario since the ACK generation time at the BS is very small. For the same reason, the waiting time for ACK will also not be very long (e.g., up to one or two downlink transmission time). Note that this scenario is quite rare and most of the times the nodes will receive ACK immediately upon transmission.

\begin{figure*}[!htbp] 
    \centering
      \subfigure[RSSI under varying distance\label{fig:csi_rssi}]{
    \includegraphics[width=0.35\textwidth]{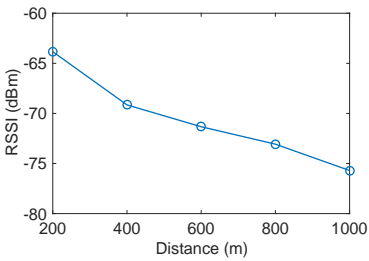}
      }\hfill
      \subfigure[Path Loss under varying distance\label{fig:csi_pathloss}]{
        \includegraphics[width=.35\textwidth]{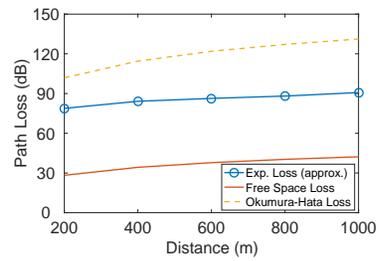}
      }\hfill
      \subfigure[BER under varying distance\label{fig:csi_ber}]{
        \includegraphics[width=.35\textwidth]{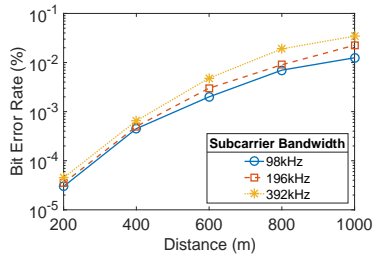}
      }
    \caption{RSSI, path loss, and BER at the SNOW BS for a TI CC1310 node.}
    \label{fig:csi}
 \end{figure*}
\subsection{Channel State Information Estimation}\label{sec:csi}

Multi-user OFDM communication requires channel estimation and tracking to ensure high data rate at the BS. One way to avoid channel estimation is to use the \emph{differential phase-shift keying (DPSK)} modulation. DPSK, however, results in a lower bitrate at the BS due to a 3dB loss in the signal-to-noise ratio (SNR)~\cite{van1995channel}. Additionally, the current SNOW design does not support DPSK modulation. SNR at the BS for each node is different in SNOW. Also, SNR of each node is affected differently due to channel conditions, deteriorating the overall bitrate in the uplinks. Thus, it requires handling of the channel estimation in SNOW.

Figure~\ref{fig:csi} shows the experimentally found received signal strength indicator (RSSI), path loss, and BER at the SNOW BS for a CC1310 device that transmits successive 1000 30-byte (payload) packets from 200 to 1000m distances, respectively, with a Tx power of 15dBm, subcarrier center frequency at 500MHz, and a bandwidth of 98kHz. Figure~\ref{fig:csi_rssi} indicates that the RSSI decreases rapidly with the increase in distance. Also, the path loss in Figure~\ref{fig:csi_pathloss} shows that it is significantly higher than the theoretical free space loss~\cite{rappaport1996wireless}. We also compare with the Okumura-Hata~\cite{rappaport1996wireless} loss to check if it fits the model, however, it does not. Finally, Figure~\ref{fig:csi_ber} confirms that the BER goes above $10^{-3}$ (which is not acceptable~\cite{rnr}) beyond 400m due to the unknown channel conditions. Figure~\ref{fig:csi_ber} also shows that the BER worsens for an increase in the subcarrier bandwidth. Thus, to make our implementation more resilient, we need to incorporate the CSI estimation in SNOW.

We calculate the CSI for each SNOW node independently on its subcarrier. We consider a slow flat-fading model~\cite{tse2005fundamentals}, where the channel conditions vary slowly with respect to a single node to BS packet duration. Note that joint-CSI estimation~\cite{jiang2007iterative, ribeiro2008uplink} in SNOW is not our design goal since it would require SNOW nodes to be strongly time-synchronized.  
Similar to IEEE 802.16e, we run CSI estimation independently for each node because of their different fading and noise characteristics. In the following, we explain the CSI estimation technique for one node on its subcarrier for each packet. The BS uses the same technique to estimate CSI for all other nodes. 

For a node, in a narrowband flat-fading subcarrier, the system is modeled as $y = Hx + w$,
where $y$, $x$, and $w$ are the receive vector, transmit vector, and noise vector, respectively. $H$ is the channel matrix. 
We model the noise as additive white Gaussian noise, i.e., a circular symmetric complex normal ($CN$) with $w \sim CN(0, W)$, where the mean is zero and noise covariance matrix $W$ is known.
As the subcarrier conditions vary, we estimate the CSI on a short-term basis based on popular approach called {\em training sequence}. We use the known preamble transmitted at the beginning of each packet. $H$ is estimated using the combined knowledge of the received and the transmitted preambles. To make the estimation robust, we divide the preamble into $n$ equal parts (preamble sequence). E.g., n = 4, which is similar to the estimation in IEEE 802.11.

Let the preamble sequence be $(p_1, p_2, \cdots, p_n)$, where vector $p_i$ is transmitted as $y_i = Hp_i + w_i$.
Combining the received preamble sequences, we get $Y = [y_1, \cdots, y_n] = HP + W$, where 
$P = [p_1, \cdots, p_n]$ and $W = [w_1, \cdots, w_n]$. With combined knowledge of $Y$ and $P$, channel matrix $H$ is estimated. Similar to the CSI estimation in the uplink communications by the BS, each node also estimates the CSI during its downlink communications. Note that the computational complexity of CSI estimation at the nodes is lightweight since each SNOW packet has a 32-bit preamble~\cite{snow_ton}, divided into four equal parts. A node thus processes a vector of only 8 bits at a time.

\subsection{Carrier Frequency Offset Estimation} \label{sec:cfo}

Multi-user OFDM systems are very sensitive to the CFO between the transmitters and the receiver. CFO causes the OFDM systems to lose orthogonality between subcarriers, which results in severe ICI. 
A transmitter and a receiver observe CFO due to (i) the mismatch in their local oscillator frequency as a result of hardware imperfections; (ii) the relative motion that causes a Doppler shift. 
ICI degrades the SNR between an OFDM transmitter and a receiver, which results in significant BER. Thus, we investigate the needs for CFO estimation in our implementation.
\begin{figure}[!htb]
\centering
\includegraphics[width=0.35\textwidth]{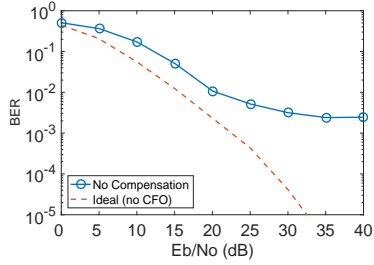}
\caption{BER at different $E_b/N_0$.}
\label{fig:cfo}
\end{figure}
The loss in SNR due to the CFO between the SNOW BS and a node can be estimated as 
$SNR_{loss} = 1 + \frac{1}{3}(\pi \delta f T)^2\frac{E_s}{N_0}$~\cite{nee2000ofdm}, where
$\delta f$ is the frequency offset, $T$ is the symbol duration, $E_s$ is the average received subcarrier energy, and $N_0/2$ is the two-sided spectral density of the noise power.

To observe the effects of CFO, we choose two neighboring orthogonal subcarriers in the BS and send concurrent packets from two nodes at 200m distance. Each node sends successive 1000 30-byte packets. We repeat this experiment varying the transmission powers at the nodes to generate signals with different $E_b/N_0$, where $E_b$ is the average energy per bit in the received signals. 
Figure~\ref{fig:cfo} shows the BER at the BS while receiving packets from these two nodes. This figure shows that BER is nearly $10^{-3}$ even for very high $E_b/N_0$ ($\approx 40$dB), which is also very high compared to the theoretical BER~\cite{choi2000carrier}. Thus, CFO is heavily pronounced in SNOW.
The distributed and asynchronous nature of SNOW does not allow CFO estimation similar to the traditional multi-user OFDM systems.
While the USRP-based SNOW implementation provides a trivial and {\em coarse} CFO estimation, it is not robust and does not account for the mobility of the nodes~\cite{snow_ton}.
We propose a pilot-based robust CFO estimation technique, combining both coarse and finer estimations, which accounts for the mobility of the nodes as well. We use training symbols for CFO estimation in an ICI free environment for each node independently, while it joins the network by communicating with the BS using a non-overlapping {\em join subcarrier}.

We explain the CFO estimation technique between a node and the BS (uplink) on a join subcarrier $f$ based on time-domain samples. Note that the BS keeps running the G-FFT on the entire BS spectrum. We thus extract the corresponding time-domain samples of the join subcarrier by applying IFFT during a node join. The join subcarrier does not overlap with other subcarriers; hence it is ICI-free. If $f_{\text{node}}$ and $f_{\text{BS}}$ are the frequencies at a node and the BS, respectively, then their frequency offset $\delta f = f_{\text{node}}-  f_{\text{BS}}$.
For transmitted signal $x(t)$ from a node, the received signal  $y(t)$ at the BS that experiences a CFO of $\delta f$ is given by 
$y(t)  = x(t) e^{j2\pi \delta f t}$.
Similar to IEEE 802.11a, we estimate $\delta f$ based on short and long preamble approach. Note that the USRP-based implementation has considered only one preamble to estimate CFO.
In our implementation, the BS first divides a $n$-bit preamble from a node into short and long preambles of lengths $n/4$ and $3n/4$, respectively. Thus for a 32-bit preamble (typically used in SNOW), the lengths of the short and long preambles are  8 and 24, respectively. 
The short preamble and the long preamble are used for coarse and finer CFO estimation, respectively. 
Considering $\delta t_s$ as the short preamble duration and $\delta f_s$ as the coarse CFO estimation, we have
$y(t-\delta t_s)  = x(t) e^{j2\pi \delta f_s (t-\delta t_s)}.$

Since $y(t)$ and $y(t-\delta t_s)$ are known at the BS, we have
\begin{align*}
y(t-\delta t_s) y^*(t)  & = x(t) e^{j2\pi \delta f_s (t-\delta t_s)}       x^*(t) e^{-j2\pi  \delta f_s t}
                           = |x(t)|^2  e^{j 2\pi  \delta f_s -\delta t_s }.
\end{align*}
Taking angle of both sides gives us as follows.
\begin{align*}
\sphericalangle  y(t-\delta t_s) y^*(t)   &=  \sphericalangle     |x(t)|^2  e^{j 2\pi  \delta f_s -\delta t_s } =      - 2\pi  \delta f_s \delta t_s
\end{align*}
By rearranging the above equation, we get
$$\delta f_s   =  - \frac{\sphericalangle  y(t-\delta t_s) y^*(t) }{2\pi\delta t_s}.$$

Now that we have the coarse CFO $\delta f_s$, we correct each time domain sample (say, $P$) received in the long preamble as $ P_a = P_a e^{-ja \delta f_s}$, where $a = \{1, 2, \cdots, A\}$ and $A$ is the number of time-domain samples in the long preamble. Taking into account the corrected samples of the long preamble and considering $\delta t_l$ as the long preamble duration, we estimate the finer CFO as follows. 
\begin{equation} 
\delta f  =  - \frac{\sphericalangle  y(t-\delta t_l) y^*(t) }{2\pi\delta t_l} \label{eqn:finer_cfo}
\end{equation}
To this extent, considering the join subcarrier $f$, the {\slshape ppm (parts per million)} on the BS's crystal is given by $ \text{ppm}_\text{BS} = 10^6  \big(\frac{\delta f}{f}\big) $. Thus, the BS calculates $ \delta f_i$ on subcarrier $f_i$ (assigned for node $i$) as 
$\delta f_i =  \frac{(f_i * \text{ppm}_\text{BS})}{10^6}.$ The CFO between the Tx-Radio and the Rx-radio can be estimated using a basic SISO CFO estimation technique~\cite{yao2005blind}. Thus, BS also knows the CFO for the downlink.

We now explain the CFO estimation to compensate for the Doppler shift. Note that if the signal bandwidth is sufficiently narrow at a given carrier frequency and mobile velocity, the Doppler shift can be approximated as a common shift across the entire signal bandwidth~\cite{talbot2007mobility}. Thus, the Doppler shift in the join subcarrier for a node also represents the Doppler shift at its assigned subcarrier, and hence the estimated CFO in Equation (\ref{eqn:finer_cfo}) is not affected due to the Doppler Shift.
For simplicity, we consider that a node's velocity is constant and the change in Doppler shift is negligible during a single packet transmission in SNOW.
Considering $\delta f_d$ as the CFO due to the Doppler shift, $v$ as the velocity of the node, and $\theta$ as the angle of the arrived signal at the BS from the node, we have $f_d = f_i\big(\frac{v}{c}\big)\cos(\theta)$~\cite{talbot2007mobility}, where
$f_i$ is the subcarrier center frequency and $c$ is the speed of light. The node itself may consider its motion as circular and approximate $\theta = \frac{\delta s}{r}$, where $\delta s$ is the amount of anticipated change in position during a packet transmission and $r$ is the {\em line-of-sight} distance between the node and BS. Thus, CFO compensation due to the Doppler shift is done at the nodes during uplink communications. In the downlink communications, the BS Tx-Radio can also compensate for the node's mobility as the node can report its Doppler shift to the BS during the uplink communications.

In summary, as the nodes asynchronously transmit, estimating joint-CFO of the subcarriers at the BS is very difficult. We thus use a simple feedback approach for proactive CFO correction in the uplink communications. Specifically, 
$\delta f_i$  estimated at the BS for subcarrier $f_i$ is given to the node (during joining process at subcarrier $f_i$).
The node may then adjust its transmitted signal based on $\delta f_i$ and $\delta f_d$, calculated as $(\delta f_i + \delta f_d)$, which will align its signal so that the BS does not need to compensate for CFO in the uplink communications. Such feedback-based proactive compensation scheme was studied before for multi-user OFDM and is also used in global system for mobile communication (GSM)~\cite{van1999time}.

\section{Handling the Near-Far Power Problem} \label{sec:near-far}
\begin{figure}[!htb]
\centering
\includegraphics[width=0.5\textwidth]{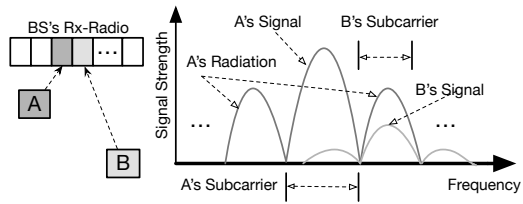}
\caption{An illustration of the near-far power problem. B is farther from the BS than A and both transmit concurrently using the same Tx power.}
\label{fig:near-far}
\end{figure}
Wireless communication is susceptible to the near-far power problem, especially in CDMA (Code Division Multiple Access)~\cite{muqattash2003cdma}. Multi-user D-OFDM system in SNOW may also suffer from this problem. Figure~\ref{fig:near-far} illustrates the near-far power problem in SNOW. Suppose, nodes A and B are operating on two adjacent subcarriers. Node A is closer to the BS compared to node B. When both nodes A and B transmit concurrently to the BS, the received frequency domain signals from node A and B may look as shown on the right of Figure~\ref{fig:near-far}. Here, transmission from node B is severely interfered by the strong radiations of node A's transmission. As such, node B's signal may be buried under node A's signal making it difficult for the BS to decode the packet from node B. 
A typical SNOW deployment may have such scenarios if the nodes operating on adjacent subcarriers use the same transmission power and transmit concurrently at the BS from different distances. 
\begin{figure}[t]
    \centering 
      \subfigure[Avg. PDR at different Tx powers\label{fig:nf_pdr}]{
    \includegraphics[width=0.35\textwidth]{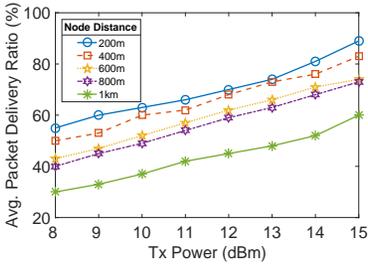}
      }\hfill
      \subfigure[Avg. PDR at different Tx powers and time\label{fig:nf_time}]{
        \includegraphics[width=.35\textwidth]{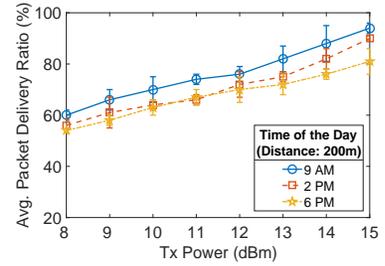}
      }
    \caption{Packet delivery ratio at different Tx powers}
    \label{fig:nf-effects}
 \end{figure}

To observe the near-far power problem in SNOW, we run experiments by choosing 3 different adjacent subcarriers, where the middle subcarrier observes the near-far power problem introduced by both subcarriers on its left and right. We place two CC1310 nodes within 20m of the BS that use the left and the right subcarrier, respectively. We use another CC1310 node that uses the middle subcarrier and is placed at different distances between 200 and 1000m from the BS. Nodes that are within 20m of the BS transmit packets continuously with a transmission power of 0dBm. At each distance, for each transmission power between 8 and 15dBm, the node that uses the middle subcarrier sends 100 rounds of 1000 consecutive packets (sends one packet then waits for the ACK and then sends another packet, and so on) to the BS and with a random interval of 0-500ms. For each transmission power level, at each distance, that node calculates its average {\em packet delivery ratio (PDR)}. PDR is defined as the ratio of the number of successfully acknowledged packets to the number of total packets sent.
We repeat the same experiments for 7 days at 9 AM, 2 PM, and 6 PM.

Figure~\ref{fig:nf_pdr} shows that the average PDR increases at each distance with the increase in the transmission power. Figure~\ref{fig:nf_time} depicts the result for 7-day experiments (only at a distance of 200m) and shows that the average PDR changes at different time of the day. Overall, Figure~\ref{fig:nf_pdr} and~\ref{fig:nf_time} confirms that the average PDR increases with the increase in the transmission power. To ensure the energy-efficiency at the nodes, i.e., to find a  transmission power  that suffices to eliminate the effects of near-far power problem, we propose an adaptive transmission power control for the SNOW design, as described below.


\subsection{Adaptive Transmission Power Control}\label{sec:atpc}
Our design objective for the adaptive Tx power control is to correlate the subcarrier-level Tx power and link quality (i.e., PDR) between each node and the BS. We thus formulate a predictive model to provide each node with a proper Tx power to make a successful transmission to the BS using its assigned subcarrier. Note that our work differs from the work in~\cite{lin2016atpc} in fundamental concepts of the network design and architecture. In~\cite{lin2016atpc}, the authors have considered a multi-hop wireless sensor network based on IEEE 802.15.4 with no concurrency between a set of transmitters and a receiver. Additionally, our model is much more simpler since we deal with single hop communications. As such, the overheads (i.e., energy consumption and latency at each node) associated with our model are fundamentally lesser than that in~\cite{lin2016atpc}, or the other techniques developed for multi-hop wireless networks~\cite{son2006experimental, li2005cone}. In the following, we describe our model.

Whenever a node is assigned a new subcarrier or observes a lower PDR, e.g., PDR below quality of service (QoS) requirements due to mobility, it runs a lightweight predictive model to determine the convenient Tx power to make successful transmissions to the BS.
Our predictive model uses an approximation function to estimate the PDR distribution at different Tx power levels. Over time, that function is modified to adapt to the node's changes. The function is built from the sample pairs of the Tx power levels and PDRs between the node and the BS via a curve-fitting approach. A node collects these samples by sending groups of packets to the BS at different Tx power levels. A node may not be assigned new subcarriers or may not observe lower PDR due to mobility (as per our CSI and CFO estimations) frequently. Thus, the overhead (e.g., energy consumption) for collecting these samples may be negligible compared to the overall network lifetime (which is several years).

Specifically, our predictive model uses two vectors: $TP$ and $L$, where $TP = \{ tp_1, tp_2, \cdots, tp_m \}$ contains $m$ different Tx power levels that the node uses to send $m$ groups of packets to the BS and $L = \{ l_1, l_2, \cdots, l_m \}$ contains the corresponding PDR values at different Tx power levels. Thus, $l_i$ represents the PDR value at Tx power level $tp_i$. We use the following linear function to correlate between Tx power and PDR.
\begin{equation}
	l(tp_i) = a~.~tp_i + b \label{eqn:linear_model}
\end{equation}
To lessen the computational overhead in the node, we adopt the {\em least square approximation} technique to determine the unknown coefficients $a$ and $b$ in Equation (\ref{eqn:linear_model}). Thus, we find the minimum of the function $S(a, b)$, where $\nonumber S(a, b) = \sum |l_i - l(tp_i)|^2.$
The minimum of $S(a, b)$ is determined by taking the partial derivatives of $S(a, b)$ with respect to $a$ and $b$, respectively, and setting them to zero. Thus, $ \frac{\partial S}{\partial a} = 0$ and $\frac{\partial S}{\partial b} = 0$ give us
\begin{align}
	\nonumber a~\sum (tp_i)^2 + b~\sum tp_i &= \sum l_i.tp_i \text{ and} \\ 
  \nonumber a~\sum tp_i + b~m &= \sum l_i.
\end{align}
Simplifying the above two equations, we find the estimated values of $a$ and $b$ as follows.
\begin{equation}\nonumber
\begin{split}
	\begin{bmatrix}
		\hat{a}\\
        \hat{b}
	\end{bmatrix}
    = \frac{1}{m \sum (tp_i)^2 - (\sum tp_i)^2} \times \\
    \begin{bmatrix}
    	m \sum l_i.tp_i - \sum l_i \sum tp_i\\
    	\sum l_i \sum (tp_i)^2 - \sum l_i.tp_i \sum tp_i
    \end{bmatrix}
\end{split}
\end{equation}
Using the estimated values of $a$ and $b$, the node can calculate the appropriate Tx power as follows.
\begin{equation}\label{eqn:estimated}
tp = \big[\frac{PDR_{\text{threshold}} - \hat{b}}{\hat{a}}\big] \in TP
\end{equation}
Here, $PDR_{\text{threshold}}$ is the threshold set empirically or according to QoS requirements, and $[.]$ denotes the function that rounds the value to the nearest integer in the vector $TP$.

Now that the initial model has been established in Equation (\ref{eqn:estimated}), this needs to be updated continuously with the node's changes over time. In Equation (\ref{eqn:linear_model}), both $a$ and $b$ are functions of time that allow the node to use the latest samples to adjust the curve-fitting model dynamically. 
It is empirically found that (Figure~\ref{fig:nf_pdr}) the slope of the curve does not change much over time; hence $a$ is assumed time-invariant in the predictive model. On the other hand, the value of $b$ changes drastically over time (Figure~\ref{fig:nf_time}). Thus, Equation (\ref{eqn:linear_model}) is rewritten as follows that characterizes the actual relationship between Tx power and PDR.
\begin{equation}
	\nonumber l(tp(t)) = a.tp(t) + b(t)
\end{equation}
Now, $b(t)$ is determined by the latest Tx power and PDR pairs using the following feedback-based control equation~\cite{lin2016atpc}.
\begin{align}
	\nonumber \Delta \hat{b}(t) &= \hat{b}(t) - \hat{b}(t+1) \\
    			\nonumber	  &= \frac{\sum^K_{k=1} [PDR_{\text{threshold}} - l_k(t - 1)]}{K} \\ 
                      &= PDR_{\text{threshold}} - l(t-1) \label{eqn:control}
\end{align}
Here, $l(t-1)$ is the average value of $K$ readings denoted as 
\begin{equation}
	\nonumber l(t-1) = \frac{\sum^K_{k=1} l_k(t - 1)}{K}.
\end{equation}
Here, $l_k(t-1)$, for $k = \{1, 2, \cdots, K\}$, is one reading of PDR during the time period $t-1$ and $K$ is the number of feedback responses at time period $t-1$. Now, the error in Equation (\ref{eqn:control}) is deducted from the previous estimation; hence the new estimation of $b(t)$ can be written as: $\hat{b}(t) = \hat{b}(t-1) - \Delta \hat{b}(t)$.
Given the newly estimated $\hat{b}(t)$, the node now can set the Tx power at time $t$ as
\begin{equation}
	\nonumber tp(t) = \big[\frac{PDR_{\text{threshold}} - \hat{b}(t)}{\hat{a}}\big].
\end{equation}

\section{Network Architecture and Deployment Cost}\label{sec:deploy}
\begin{figure}[!htb]
\centering
\includegraphics[width=0.5\textwidth]{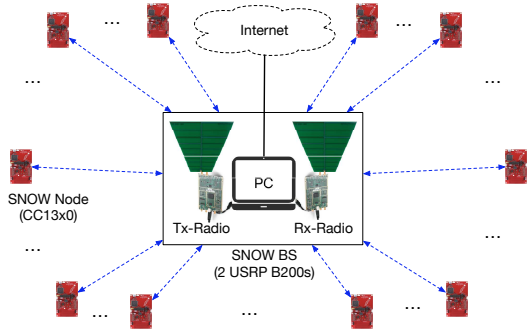}
\caption{The SNOW architecture for practical deployment (The PC may be replaced by a Raspberry Pi device. The two USRP B200 devices can be replaced by a USRP B2100 device that has two half-duplex radios.)}
\label{fig:deployment}
\end{figure}
In this section, we discuss the practical applicability of our implementation. Figure~\ref{fig:deployment} shows our network view. The SNOW BS is a PC that connects two USRP B200 devices (Tx-Radio and Rx-Radio). The BS is also connected to the Internet. In the BS, a USRP B210 device may be used which has two half-duplex radios. Also, a Raspberry Pi device may be used instead of the PC. All the CC13x0 nodes are battery-powered and directly connected to the BS. 
\begin{figure}[!htb]
\centering
\includegraphics[width=0.35\textwidth]{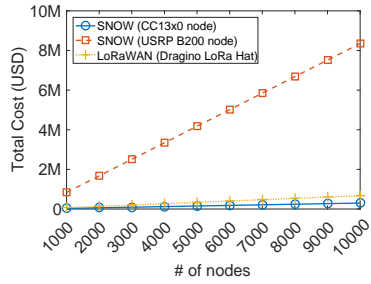}
\caption{Practical deployment cost with numerous nodes.}
\label{fig:cost}
\end{figure}

We now analyze the deployment cost of our CC13x0-based SNOW implementation and compare with the original USRP-based SNOW implementation in~\cite{snow_ton}. 
Figure~\ref{fig:cost} shows the total deployment cost of our CC13x0-based SNOW implementation for different numbers of nodes between 1000 and 10,000. A CC1310 or CC1350 device costs approximately \$30 USD (retail price). The price for the BS is approximately \$1600 USD (two USRP B200 devices \$750 USD each, and two antennas \$50 USD each). In this comparison, the cost of the PC is not considered since it is common for both implementations. For SNOW implementation in~\cite{snow_ton}, a node is a  USRP B200 device that has an antenna and runs on a Raspberry Pi. A Raspberry Pi device costs approximately \$35 USD.

To provide an insight into the deployment cost of a LoRaWAN network, we consider the Dragino LoRa/GPS-Hat (e.g., SX1276 chip) that runs on Raspberry Pi and costs approximately \$32 USD (retail price)~\cite{dragino}. We choose this LoRaWAN node since it has almost identical computational and RF capabilities, compared to the TI CC13x0 devices (e.g., both have the Cortex-M series MCU, similar energy profiles, same set of sensors, and software support). In addition, we consider a LoRaWAN gateway that costs approximately \$299 USD and can receive packets on multiple channels simultaneously~\cite{loracost}. We rule out cheaper LoRaWAN devices (costs $\approx$\$10 USD) from the calculation since they do not have a similar profile as CC13x0 and do not provide software support.

As shown in Figure~\ref{fig:cost}, to deploy an LPWAN with 1000 nodes, the CC13x0-based SNOW implementation may cost approximately \$31.6K USD, compared to \$836.6K USD for the USRP-based SNOW implementation proposed in~\cite{snow_ton}, and \$67.3K USD for the Dragino LoRa-Hat-based LoRaWAN. 
For a deployment of 10,000 nodes, the costs are \$301.6K, \$8.3M, and \$670.3K USD
for CC13x0-based SNOW implementation, USRP-based SNOW implementation, and Dragino LoRa-Hat-based LoRaWAN, 
respectively. As shown in Figure~\ref{fig:cost}, the cost of each LPWAN increases linearly with the increase in the number of nodes. However, the cost of our CC13x0-based SNOW implementation in unnoticeable.
Our new implementation of SNOW on the CC13x0 devices thus becomes highly scalable in terms of cost, making SNOW deployable for practical applications.

\section{Evaluation}\label{sec:eval}
In this section, we provide an extensive evaluation (both uplink and downlink performances) of our CC13x0-based SNOW implementation, considering both stationary and mobile CC13x0 nodes.

\subsection{Setup}\label{sec:expsetup}
\begin{figure}[!htb]
\centering
\includegraphics[width=0.35\textwidth]{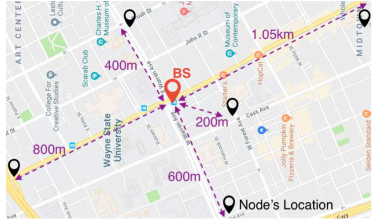}
\caption{SNOW deployment in Detroit, Michigan.}
\label{fig:testbed}
\end{figure}

Figure~\ref{fig:testbed} shows our deployment in the city of Detroit, Michigan. Specifically, we deploy 22 CC1310 devices and 3 CC1350 devices (25 CC13x0 devices in total).
Due to our limited resources (e.g., batteries/power supply options outdoors), we create five (denoted by black pins) different clusters from 25 nodes at distances 200, 400, 600, 800, and 1000m from the BS (denoted by orange pin as well as a label). In each cluster, we randomly place 5 CC1310 devices or a mixture of CC1310 and CC1350 devices that are connected to a laptop via a USB hub so that they can operate without interruptions.
In our deployment, one group of nodes may be hidden from another group of nodes, thus creating hidden terminal scenarios. For example, the nodes at 1.05km and 800m (see Figure~\ref{fig:testbed}) may be hidden from each other since they are placed in the opposite directions of the BS.
We use a white space channel with frequency band 500--506MHz and split into 29 (numbered 1--29) overlapping (50\%) orthogonal subcarriers, each 400kHz wide. The USRP-based SNOW also uses a similar subcarrier bandwidth~\cite{snow_ton}.
We use the 28th subcarrier as the join subcarrier and the 26th subcarrier as the downlink subcarrier. We do not use the 29th and the 27th subcarriers so that the join subcarrier may remain ICI-free.
The remaining 25 subcarriers are assigned to different nodes.

We use the packet structure of the CC13x0 devices (\code{preamble}: (32 bits), \code{sync word}: (32 bits), \code{paylod length}:, \code{payload}: variable length, and \code{CRC} (16 bits)). Our default payload length is 30 bytes and contains random data. Our default bandwidth at the CC13x0 nodes is 39kHz. We use OOK modulation supported by the CC13x0 devices. Unlike the USRP-based SNOW, we do not use any spreading factor. Since the subcarrier bandwidths at the BS and the CC13x0 nodes are 400kHz and 39kHz, respectively, the oversampling at each subcarrier in the BS compensates for the spreading factor. Our default transmission power at the BS and the nodes is 15dBm. However, a CC13x0 device may choose to operate with any transmission power between 0 and 15dBm, as needed by our ATPC model. The receive sensitivity at the BS is set to -114dBm, as per the white space regulations~\cite{whitespaceSurvey}. Unless stated otherwise, these are the default parameter settings.

\begin{figure*}[!htbp]
    \centering
      \subfigure[Packet reception rate vs. distances\label{fig:prr_dist}]{
    \includegraphics[width=0.35\textwidth]{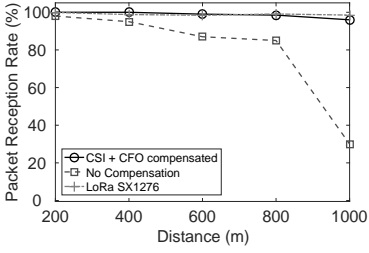}
      }\hfill
      \subfigure[Packet reception rate in uplink\label{fig:prr_txs}]{
        \includegraphics[width=.35\textwidth]{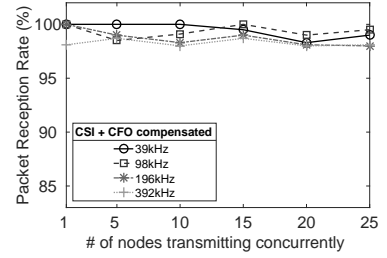}
      }\hfill
      \subfigure[Packet reception rate in downlink\label{fig:down_prr}]{
        \includegraphics[width=0.35\textwidth]{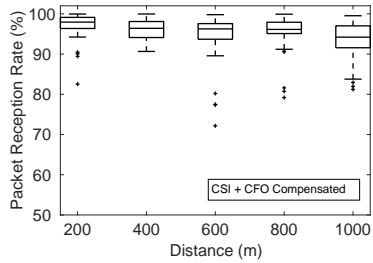}
      }
    \caption{Reliability in long distance communication.}
    \label{fig:prr}
\end{figure*}
\subsection{Reliability over Long Distance}\label{sec:prr_dist}
\subsubsection{Achievable Distance}
We first test the achievable communication range of CC13x0-based SNOW. We take one CC13x0 device and transmit to the BS from different distances between 200 and 1000m. We keep our antenna height at 3 meters above the ground for both the BS and node. At each distance, each CC13x0 transmits 1000 packets with a random interval between 0 and 500ms. The transmission power is set to 15dBm. To show comparison, we repeat the same experiments without compensating for CSI and CFO as well. Additionally, we test the achievable distance between two LoRa SX1276 devices (bandwidth: 125kHz, spreading factor: 12, coding rate: 4/5) with the above settings. We choose a bandwidth of 125kHz in LoRa since it is the closest value compared to the bandwidth in CC13x0 (39kHz).

Figure~\ref{fig:prr_dist} shows that the {\em packet reception rate (PRR)} at the SNOW BS when packets are sent with and without compensating for CSI and CFO, comparing with LoRa. PRR is defined as the ratio of the number of successfully decoded packets to the number of total packets sent.
As shown in this figure, with CSI and CFO compensation, the BS achieves 95\% of PRR from a distance of 1km . 
Without CSI and CFO compensation, the PRR at the BS is as low as 30\% from 1km distance. This figure also shows that a LoRa SX1276 device can deliver packets to another over 1km with a PRR of 96\%, which is similar to the CC13x0-based SNOW node (CSI and CFO compensated). The results thus demonstrate that SNOW on the new platform is highly competitive against LoRa, an LPWAN leader that operates in the ISM band. Additionally,
we find that beyond approximately 1km, PRR stars decreasing in our implementation. Our best guess is that if we can place the BS or the node at a higher altitude (FCC allows up to 30 meters), we may achieve high reliability over much longer communication range.

\subsubsection{Uplink Reliability}
To show the uplink reliability under concurrent transmissions from different numbers of nodes (CFO and CSI compensated), we transmit from 1, 5, 10, 15, 20, and 25 nodes to the BS, respectively. In this experiment, all the nodes are distributed within 200 and 1000m of the BS, as shown in Figure~\ref{fig:testbed}. Each node uses different subcarrier bandwidths between 39 and 392kHz. For each bandwidth starting from 39kHz, a node sends consecutive 1000 packets. Between each bandwidth, a node sleeps for 500ms. Thus, the BS knows the change in the bandwidth. 
Note that in practical deployment scenarios, a node can let know the BS of its bandwidth during the joining process.
In this experiment, we show the performance of a node for different bandwidths. Figure~\ref{fig:prr_txs} shows that we can achieve up to 99\% PRR when 25 nodes transmit concurrently using 39kHz, and up to 98.1\% PRR using 392kHz. Additionally, the number of concurrent transmissions does not affect the trend in PRR for any given bandwidth.
Thus, ensuring high uplink reliability of our CC13x0-based implementation over long distances.

\subsubsection{Downlink Reliability}
In the downlink, we test the reliability by sending 100 consecutive 30-byte (payload length) packets to each of the 25 nodes that are distributed within 200 and 1000m of the BS. We repeat the same experiment 50 times with an interval between 0 and 500ms. In this experiment, we compensate for both CSI and CFO.
Figure~\ref{fig:down_prr} shows our downlink reliability at different distances observed by different nodes. For better representation, we cluster the nodes that are located approximately at the same distance and plot the PRR against distance.
As shown in this figure, the PRR in the downlink is as high as 99\% for 75\% of the nodes that are approximately 200m away from the BS. Also, 75\% of the nodes that are approximately 1km away from the BS achieve a PRR of 96\%. Thus, this experiment confirms high downlink reliability of our CC13x0-based implementation over long distances.

\subsection{Performance in Uplink Communication}\label{sec:tputuplink}
 
In this section, we evaluate the uplink network performance in terms of throughput (achievable bitrate), end-to-end delay (time between the transmission and ACK reception of a packet), and energy consumption. We allow from 1, 5, 10, 15, 20, and 25 nodes to transmit concurrently to the BS, respectively. We distribute the nodes between 200 and 1000m in our testbed. Each node transmits 1000 30-byte (payload length) packets with a random packet interval between 0 and 100ms. 
Such packet interval confirms that the node's transmissions are indeed asynchronous to the BS.
Each node uses a bandwidth of 39kHz. 
We evaluate the uplink network performance for three different cases: (1) nodes or/and BS {\bf compensate} for CSI, CFO, and ATPC; (2) nodes or/and BS {\bf compensate only} for CSI and CFO, but not ATPC; (3) nodes or/and BS {\bf do not compensate} for CSI, CFO, and ATPC. Note that ATPC applies to the nodes only, and hence we use "or/and" in the above three cases.
For each case, we run the experiments as long as at least 90\% of the packets are delivered to the BS. A node may thus try several retransmissions.
\begin{figure*}[!htbp]
    \centering
      \subfigure[Throughput\label{fig:tput}]{
    \includegraphics[width=0.35\textwidth]{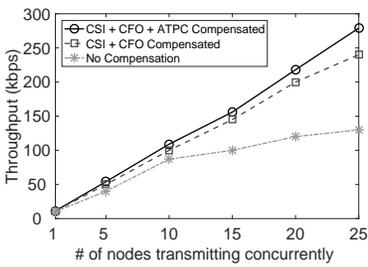}
      }\hfill
      \subfigure[End-to-end delay\label{fig:delay}]{
        \includegraphics[width=.35\textwidth]{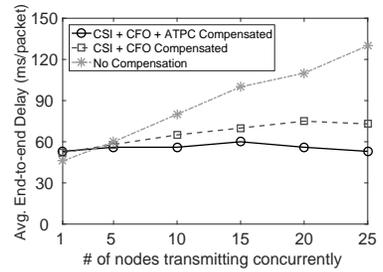}
      }\hfill 
      \subfigure[Energy consumption\label{fig:energy}]{
        \includegraphics[width=.35\textwidth]{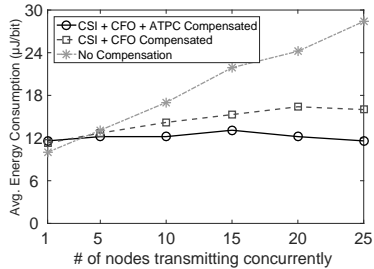}
      }
    \caption{Network performance in uplink under varying number of nodes.}
    \label{fig:performance}
 \end{figure*}

\subsubsection{Throughput}
Figure~\ref{fig:tput} shows that the BS achieves 279kbps of throughput when 25 nodes transmit concurrently (case 1), yielding approximately 11.16kbps per node. Additionally, the overall throughput at the BS increases linearly with an increase in the number of nodes transmitting concurrently. When only CSI and CFO are compensated for, the overall throughput at the BS also increases with an increase in the number of concurrent transmissions, however, it depends on the nodes' distribution (physical) across the network. If there is no near-far power problem, the overall throughput may be the same as observed in case 1. With no compensation, the throughput per node is approximately 5kbps, thus more than 2x lesser than case 1. Note that a CC13x0 device can generate a baseband signal with a symbol rate of 11.2kbaud (OOK modulated). Thus, using a node bandwidth of 39kHz or 392kHz will not affect the per node throughput. However, a lower node bandwidth gives higher PRR (Section~\ref{sec:prr_dist}) due to longer symbol duration, combating the ICI to some extent.
Additionally, if we use any other COTS device that can generate a higher symbol rate for OOK at higher node bandwidth, the per node throughput may also increase with an increase in the node bandwidth. Similarly, adopting a quadrature amplitude modulation (QAM) or frequency modulation (FM) (both yet to be explored in SNOW) at the nodes may increase the throughput at the SNOW BS. In the future, we shall adopt QAM-based or FM-based modulations in SNOW.
Overall, our CC13x0-based SNOW implementation shows high potential for practical deployments of low-rate IoT applications~\cite{whitespaceSurvey, ismail2018low}.

\subsubsection{End-to-end Delay}
Figure~\ref{fig:delay} shows the average end-to-end delay per packet at the nodes. When CSI, CFO, and ATPC are compensated for, the average end-to-end delay per packet in the network is 53ms with 25 concurrent transmissions. Also, for case 1, the average end-to-end delay per packet almost remains constant for any number of concurrent transmissions. For case 2, where only CSI and CFO are compensated for, the average end-to-end delay per packet increases a little bit with an increase in the number of concurrent transmissions. With no compensation, the average end-to-end delay per packet increases almost linearly with an increase in the number of concurrent transmissions. The reason is that a node retransmits several packets several times.  
Overall, our CC13x0-based SNOW implementation shows great promise for low-latency Industry 4.0 applications and their deployments~\cite{modekurthy2018utilization}.

\subsubsection{Energy Consumption}
Figure~\ref{fig:energy} shows the average energy consumption per bit at the nodes. We use the CC13x0 energy profile to calculate the energy consumption during Tx, Rx, and idle time~\cite{cc1350}.
For case 1, where the CSI, CFO, and ATPC are compensated for, the average energy consumption per bit in the network is approximately 11.6$\mu$J with 25 concurrent transmissions. Also, the average energy consumption per bit almost remains constant for any number of concurrent transmissions. For case 2, where only CSI and CFO are compensated for, the average energy consumption per bit increases to 16$\mu$J for 25 concurrent transmissions. Also, when nothing is compensated for, the average energy consumption per packet increases almost linearly with an increase in the number of concurrent transmissions. The reason is that a node retransmits several packets several times. Overall, small energy consumption in case 1 confirms that our CC13x0-based SNOW implementation may host long-lasting deployments.


\subsection{Performance in Downlink Communication}
\begin{figure}[!htb]
\centering
\includegraphics[width=0.35\textwidth]{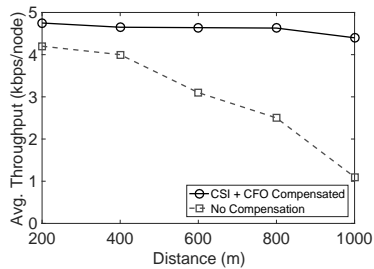}
\caption{Average throughput per node in downlink.}
\label{fig:down_tput}
\end{figure}
In this section, we evaluate the downlink network performance in terms of throughput. The BS sends 1000 consecutive 30-byte (payload length) packets to each of the 25 nodes. Also, the BS and the nodes compensate for both CSI and CFO. In the downlink, the BS uses a Tx bandwidth of 39kHz. We repeat the above experiment without compensating for CSI and CFO as well. Figure~\ref{fig:down_tput} shows the average throughput per node at different distances. For better representation, we cluster the nodes that are located approximately at the same distance and plot average throughput against the distance. As shown in this figure, a node that is approximately 200m away from the BS can achieve an average downlink throughput of 4.75kpbs, while both the BS and the node compensate for CSI and CFO. The average throughput remains almost the same as those observed at other distances, up to 1km as well. In contrast, the average throughput drops sharply with an increase in the distance when CSI and CFO are not compensated for. Note that a CC13x0 device can successfully receive an OOK-modulated signal with 4.8kbaud symbol rate and 39kHz bandwidth~\cite{snow_cots}.
Overall, our CC13x0-based SNOW implementation holds the potentials for low-rate IoT applications.

\subsection{Performance under Mobility}\label{sec:mobility_per}
In this section, we evaluate the network performance under CC13x0 node's mobility in terms of throughput, energy consumption, and end-to-end delay. 
We allow all 25 nodes to transmit concurrently to the BS. However, due to our limited resources, we enable mobility in only one node that is approximately 600m far from the BS and calculate its performance.
All nodes except the mobile node continuously transmit to the BS 30-byte (payload size) packets with a random interval between 0 and 50ms, using their assigned subcarriers, each 39kHz wide. Our CFO estimation technique (Section~\ref{sec:cfo}) allows a mobile node to travel in any direction but at a uniform speed during a packet transmission.
We vary the speed of the mobile node for different packets to 5mph, 10mph, and 20mph in any arbitrary direction within our network range. At each speed, we change the payload size of the mobile node between 10 and 120bytes. For each payload size, the mobile node transmits to the BS 1000 packets with a random interval between 0 and 50ms. We run experiments with the above settings for two cases: (1) the mobile node or/and the BS {\bf compensate} for CSI, CFO, and ATPC; (2) the mobile node or/and the BS {\bf do not compensate} for CSI, CFO, and ATPC.
\begin{figure}[t]
    \centering
      \subfigure[Throughput \label{fig:m_tput}]{
    \includegraphics[width=0.35\textwidth]{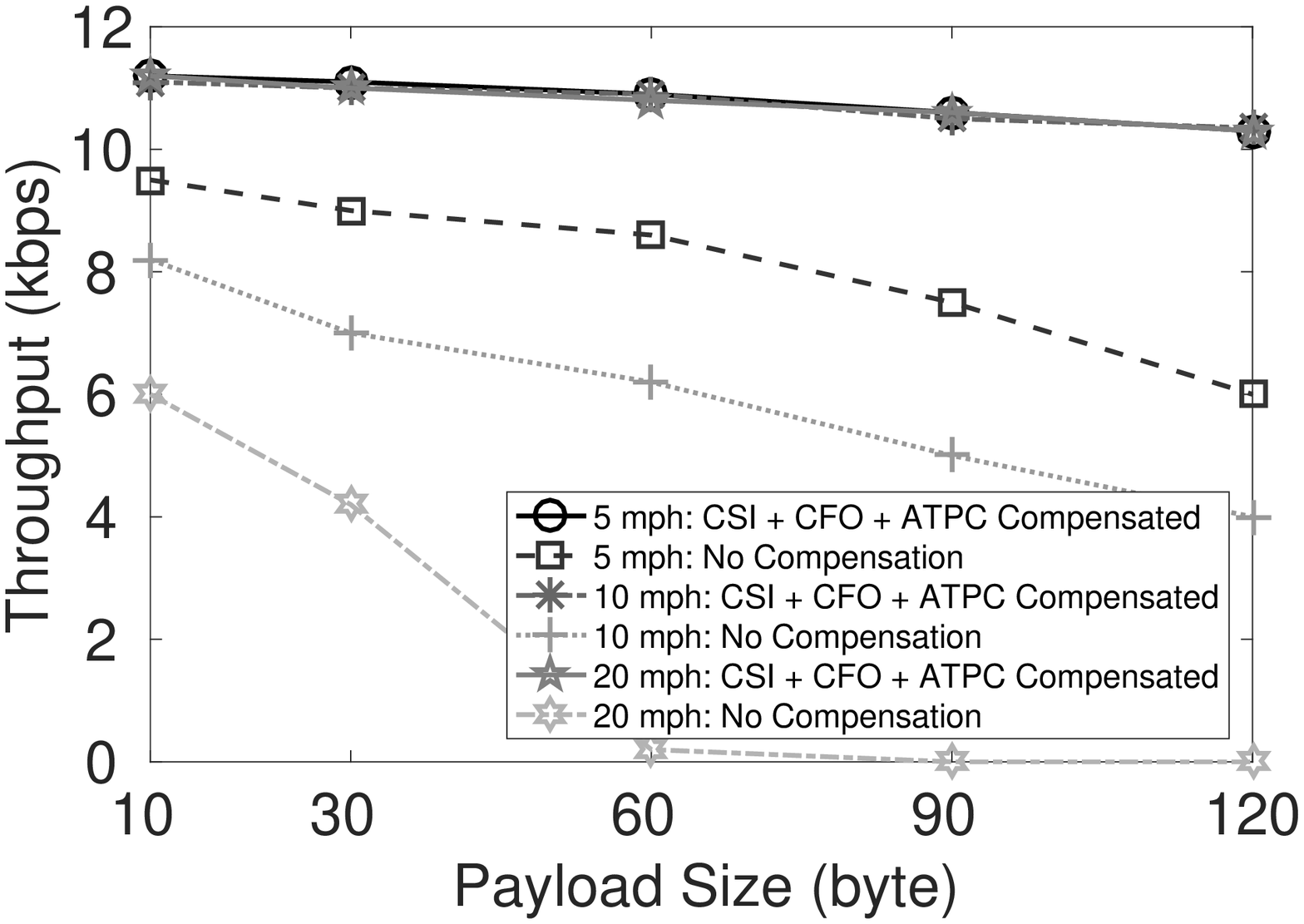} 
      }\hfill
      \subfigure[Energy consumption \label{fig:menergy}]{
        \includegraphics[width=.35\textwidth]{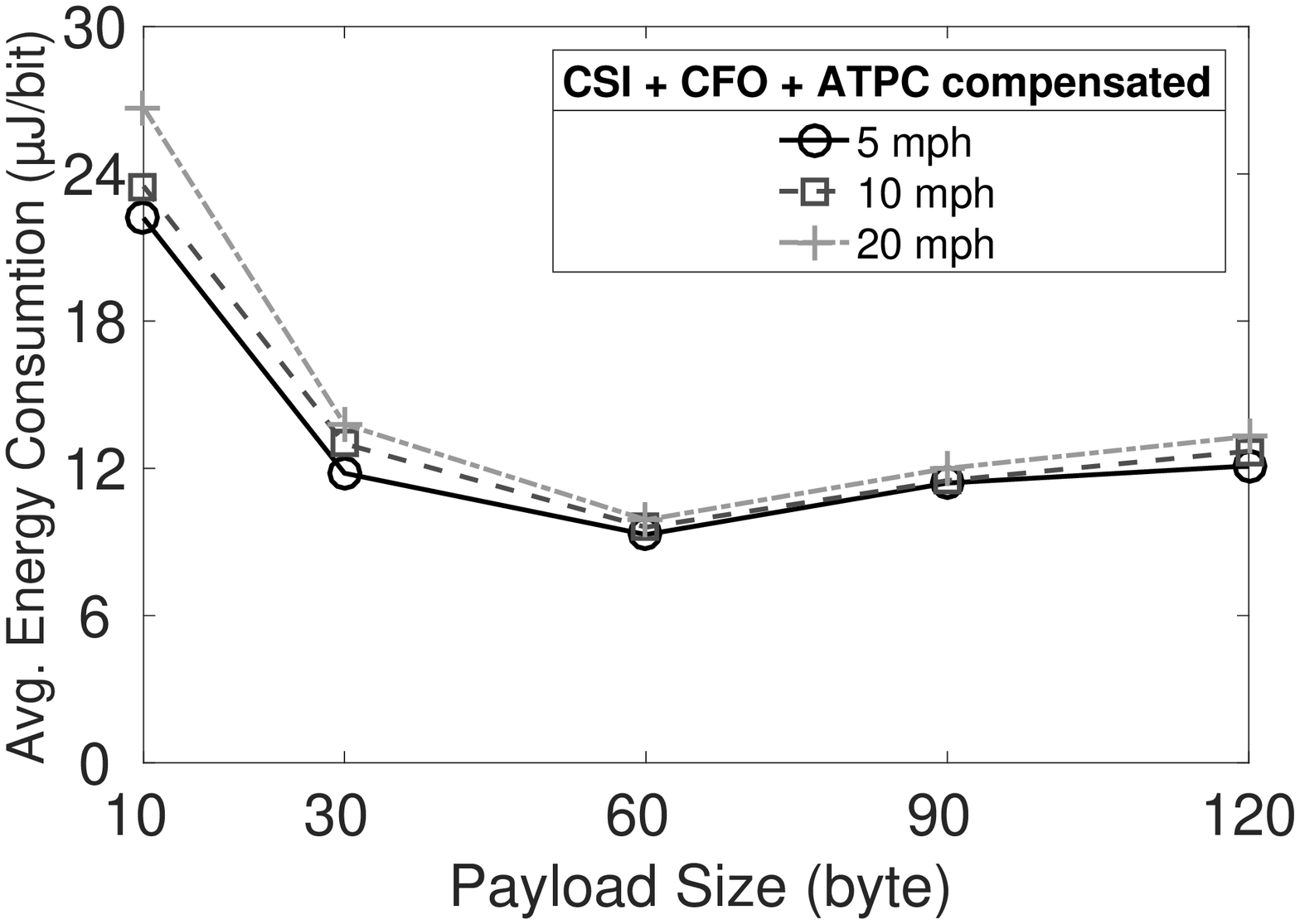}
      }
    \caption{Throughput and energy consumption under mobility.}
    \label{fig:mobility}
\end{figure}

\subsubsection{Throughput}
Figure~\ref{fig:m_tput} shows the throughput at the BS (of the mobile node) for different speeds and payload sizes. As this figure suggests, the throughput decreases slightly from 11.18kbps to 10.3kbps at 5mph, 10.35kbps at 10mph, and 10.3kbps at 20mph for an increase in the payload size between 10 and 120bytes, as CSI, CFO, and ATPC are compensated for. When the mobile node or/and the BS do not compensate for CSI, CFO, and ATPC, the throughput decreases sharply with an increase in speed and packet size. For example, at 20mph, the throughput drops to approximately 0 for payload size of 60bytes. In general, the packet size is susceptible to node's mobility. In fact, if CSI and CFO are not compensated for, the effects of unknown channel conditions and frequency offset ripple through a longer packet and increase the BER. Thus, our SNOW implementation is resilient and robust under the mobility of the nodes.

\subsubsection{Energy Consumption}

Figure~\ref{fig:menergy} shows an interesting behavior of the energy consumption per bit of the CC13x0 devices. At each speed (CSI, CFO, or/and ATPC compensated), the average energy consumption per bit decreases when we increase the payload size from 10 to 60bytes. But, when we increase the payload size beyond 60bytes (e.g., 90 and 120bytes), the energy consumption per bit starts to increase. For example, the average energy consumption per bit (at 5mph) for payload sizes 10, 30, 60, 90, and 120bytes are 22.2, 11.8, 9.3, 11.2, and 12.1$\mu$J, respectively. At speeds 10 and 20mph, the trends in energy consumption per bit are also similar to the above trend.
To the best of our knowledge, CC13x0 devices show such behavior due to their arbitrary end-to-end delays of packets with different payload lengths (to be discussed in Section~\ref{sec:e2edspeed}). Overall,  a payload of length 60bytes may be preferable (in terms of energy per bit) in the CC13x0-based SNOW implementation.


\subsubsection{End-to-end Delay}\label{sec:e2edspeed}
\begin{figure*}[t]
    \centering
      \subfigure[End-to-end delay\label{fig:mdelay}]{
        \includegraphics[width=.35\textwidth]{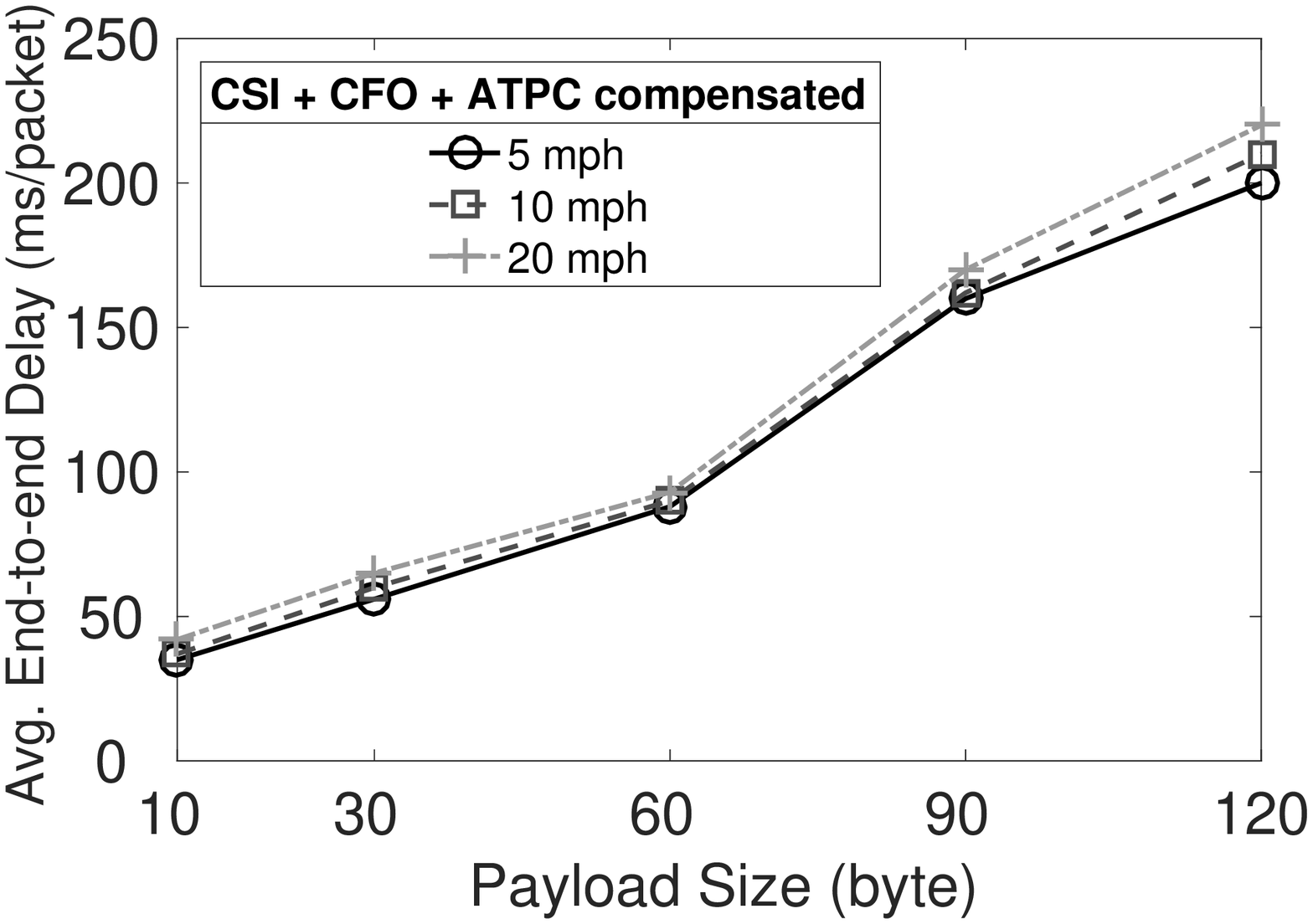}
      }\hfill 
      \subfigure[CDF of end-to-end delay\label{fig:m_cdfpayload}]{
        \includegraphics[width=0.35\textwidth]{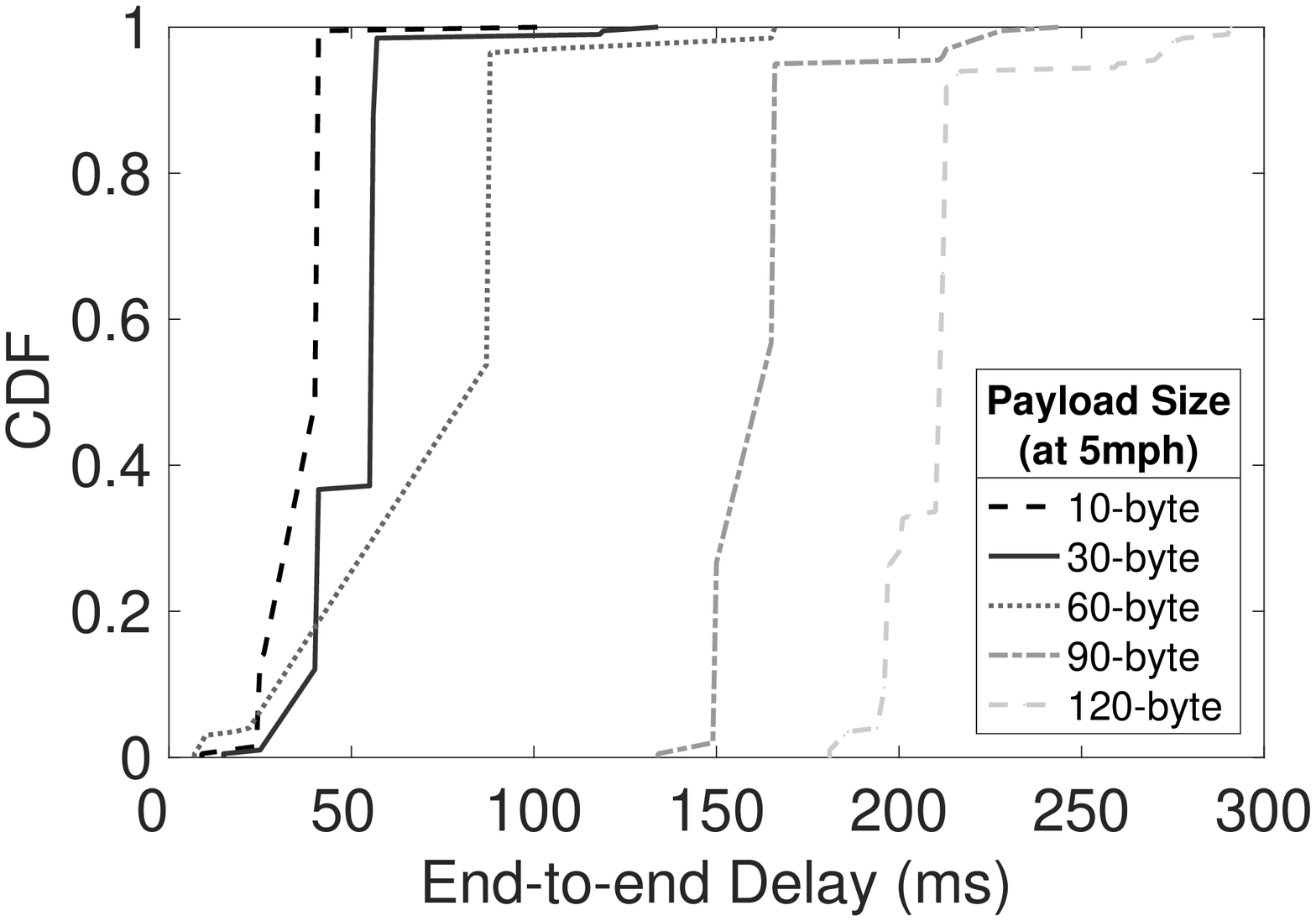}
      }\hfill
      \subfigure[CDF of end-to-end delay\label{fig:m_cdfmph}]{
        \includegraphics[width=.35\textwidth]{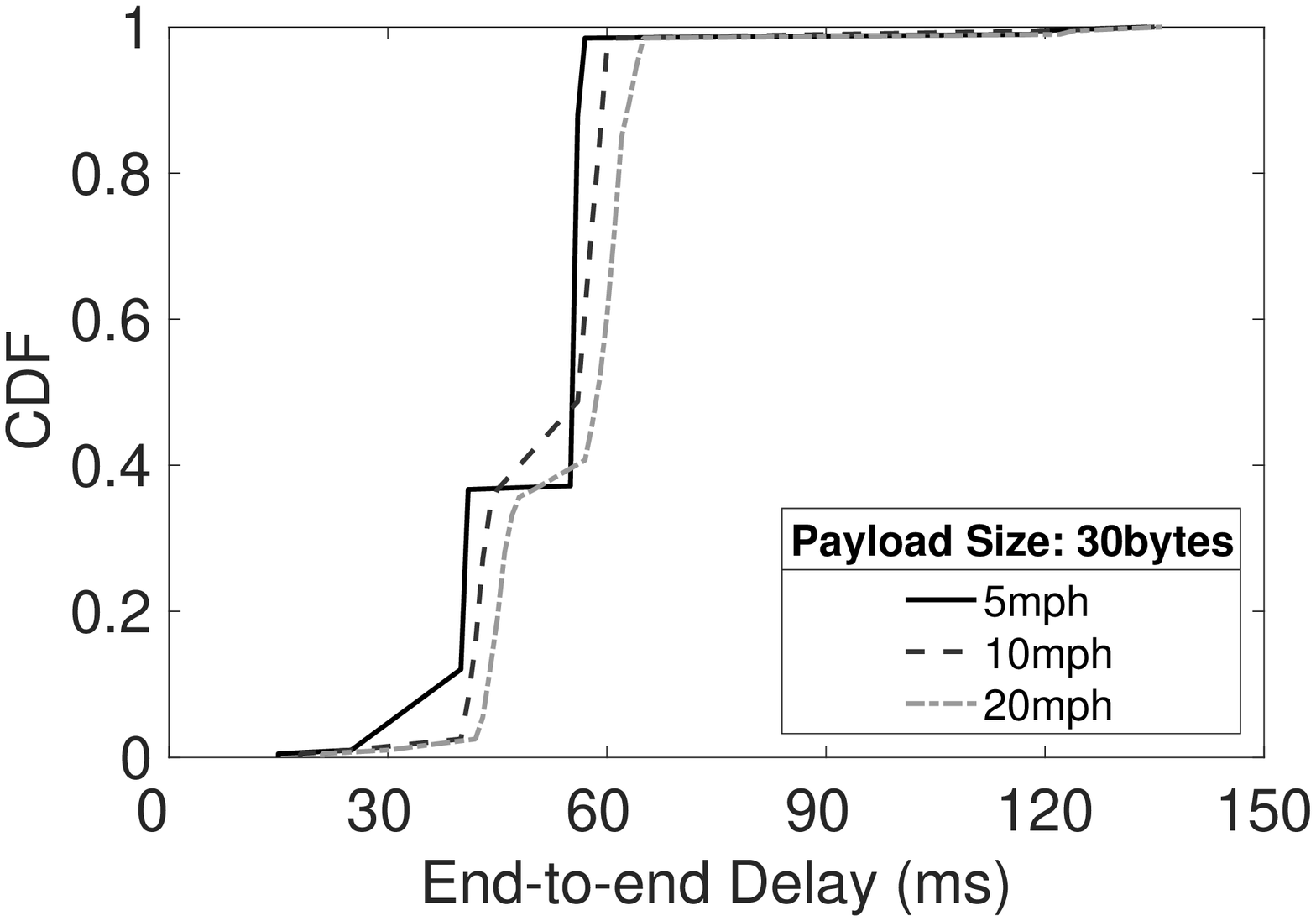}
      }
    \caption{End-to-end delay analysis of mobile node under varying payload length.}
    \label{fig:m_eted}
\end{figure*}

Figure~\ref{fig:mdelay} shows that the average per-packet end-to-end delay at the mobile node increases with an increase in speed and payload size. For example, at 5mph, the average per-packet end-to-end delays with payloads of sizes 10, 30, 60, 90, and 120bytes are 35, 56, 88, 160, 200ms, respectively; at 20mph, the average end-to-end delays are 42, 65, 93, 170, 220ms, respectively. Note that these delays in terms of payload lengths are in fact arbitrary. For example, the ratio between the delays of 30-byte and 10-byte payloads is not 30/10=3x, but 56/35=1.6x at 5mph and 65/42$\approx$1.5x at 20mph speed at the node.

Figure~\ref{fig:m_cdfpayload} shows the cumulative distribution function (CDF) of the end-to-end delay at a constant speed of 5mph with varying payload sizes. This figure shows that 60\% of the 10-byte (payload length) packets observe an end-to-end delay more than 35ms, 65\% of the 30-byte (payload length) packets observe an end-to-end delay more than 55ms, 50\% of the 60-byte (payload length) packets observe an end-to-end delay more than 90ms, 98\% of the 90-byte (payload length) packets observe an end-to-end delay more than 150ms, and 95\% of the 120-byte (payload length) packets observe an end-to-end delay more than 195ms. 
Furthermore, Figure~\ref{fig:m_cdfmph} shows the CDF of end-to-end delays for a fixed payload length of 30bytes at varying speed. As this figure shows, 98\% of the packets at 5mph observe an end-to-end delay up to 55ms, 99.99\% of the packets at 10mph observe an end-to-end delay up to 60ms, and 98\% of the packets at 20mph observe an end-to-end delay up to 65ms. Overall, Figure~\ref{fig:m_eted} confirms that our CC13x0-based SNOW implementation may provide very low latency under the mobility of the nodes.
\begin{figure}[!htb]
\centering
\includegraphics[width=0.35\textwidth]{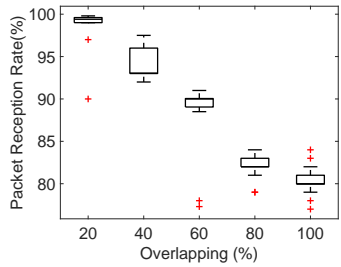}
\caption{Performance under interference.}
\label{fig:ws_interf}
\end{figure}
\subsection{Performance in the Presence of Interference}
In this section, we evaluate the performance of our implementation in the presence of interference. Due to the COVID-19 restrictions, we limit this experiment in an indoor area of (20$x$30)m$^2$.
We place a USRP B200 device within 10m of the BS to act as an interferer. Within 200ms intervals, the interferer randomly operates near the 25 subcarriers used by the CC13x0 devices and transmits random 40-byte payloads. The CC13x0 devices send packets to the BS concurrently and incessantly 
from distances between 20 and 30m. 
We let the interferer overlap 20\%, 40\%, 60\%, 80\%, and 100\% with the legitimate subcarriers. For each magnitude of overlaps, we run this experiment for 2 minutes. In Figure~\ref{fig:ws_interf}, we show the distribution of PRR at the BS (CSI, CFO, and ATPC are compensated) for 30 runs of the above experiment.
As shown in this figure, even with 100\% overlap, the PRR at the BS can be as high as 84\%. For 60\% overlap, the PRR at the BS is as high as 92\%. Overall, as we decrease the percentage of overlaps, the PRR increases at the BS. This experiments thus confirms that the impact of external interference is less severe or negligible when the interferer's spectrum partially overlaps with the legitimate subcarriers in our CC13x0-based SNOW implementation.


\subsection{Performance Comparison with LoRaWAN}

In this section, we experimentally compare the performance of our CC13x0-based SNOW implementation with a LoRaWAN network. We have 8 Dragino LoRa/GPS-Hat Sx1276 transceivers that can transmit or receive on a single channel. We create a LoRaWAN gateway capable of receiving on 3 channels simultaneously using 3 of our LoRa-Hats, while the remaining 5 devices act as LoRaWAN nodes. For a fair comparison, we allow 5 SNOW nodes (3 CC1350 devices and 2 CC1310 devices) to transmit to the SNOW BS, allowing only 3 subcarriers for data Rx/Tx. Similarly, in LoRaWAN, 5 nodes transmit on three 500kHz channels using a spreading factor of 7 and a coding rate of $\frac{4}{5}$.
In SNOW, the nodes use a subcarrier bandwidth of 392kHz with no bit spreading factor. While choosing 500kHz or 392kHz has no differentiable impact in our CC13x0-based SNOW implementation (as discussed in Section~\ref{sec:tputuplink}), we choose the latter due to the configurable Tx bandwidth limitation of the devices. The LoRaWAN gateway uses 3 adjacent 500kHz channels in the 915MHz frequency band (in the US), while the SNOW BS, in this setup, uses 3 adjacent overlapping subcarriers, numbered 10, 11, and 12 (please refer to Section~\ref{sec:expsetup} for subcarrier allocation).

The above configuration for LoRaWAN will result in its best possible throughput and energy-efficiency~\cite{snow_ton}. Additionally,  choosing LoRaWAN's largest spreading factor (e.g., 12) and 125kHz channel bandwidth for better reliability will make the comparison unfair with SNOW (e.g., SNOW uses 392kHz bandwidth and no spreading factor).
Each node (for both LoRaWAN and SNOW) transmits 1000 thirty-byte (payload size) packets from a distance of approximately 1km to the gateway/BS with a random inter-packet interval between 500 and 1000ms and a Tx power of 15dBm. Each node randomly hops to a different channel/subcarrier after sending 200 packets. In LoRaWAN, the nodes use the pure ALOHA MAC protocol, and thus operating as the Class-A LoRaWAN nodes~\cite{ismail2018low}. In SNOW, the nodes use the lightweight CSMA/CA MAC protocol (as discussed in Section~\ref{sec:snow_overview}).
In the following, we compare LoRaWAN and SNOW in terms of reliability, throughput, and energy consumption with the above settings. Note that achievable distance comparison between SNOW and LoRaWAN has already been presented in Section~\ref{sec:prr_dist}.

\begin{figure*}[t]
    \centering
      \subfigure[Packet reception rate at the gateway/BS\label{fig:lora_prr}]{
    \includegraphics[width=0.35\textwidth]{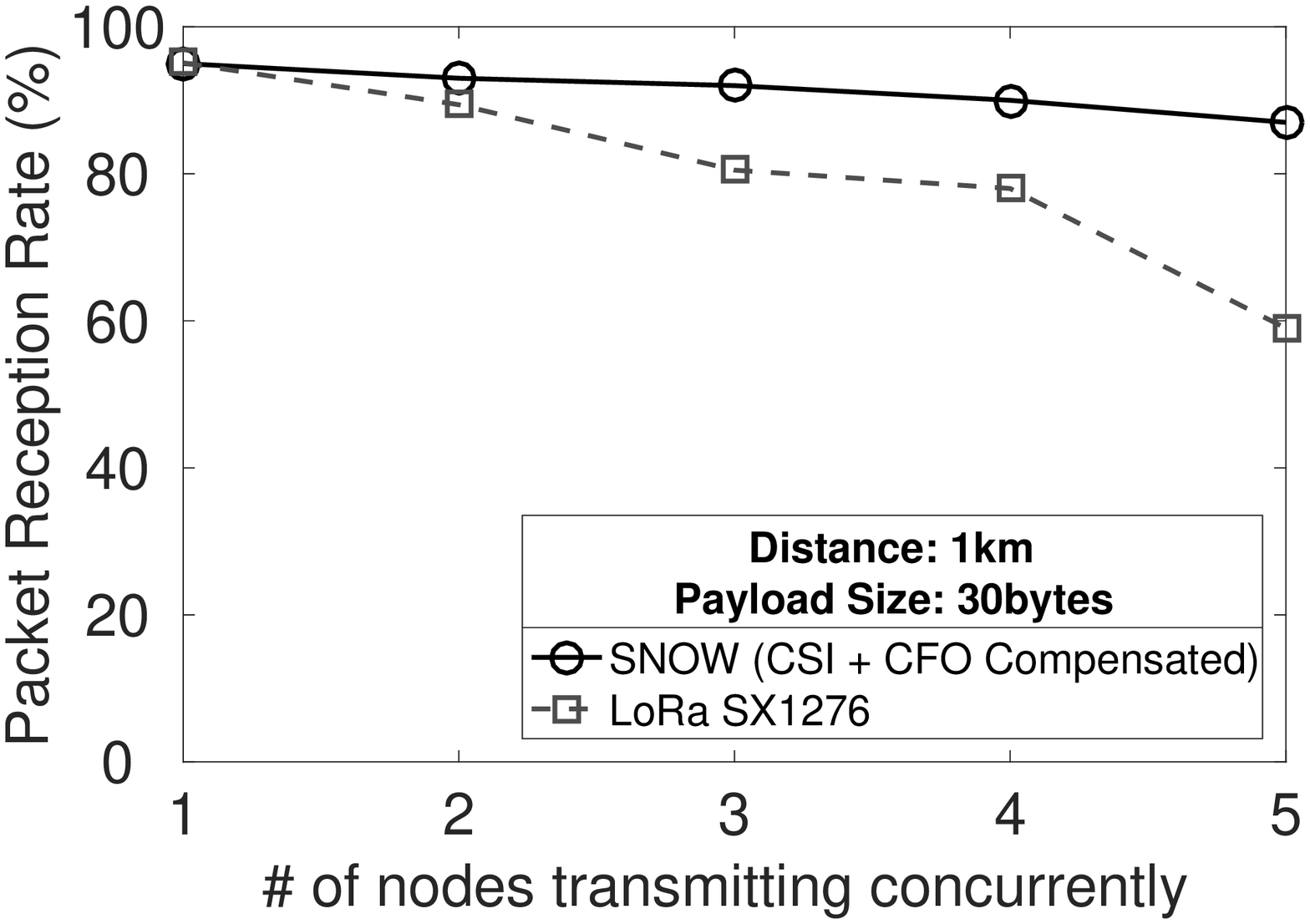}
      }\hfill
      \subfigure[Throughput at the gateway/BS\label{fig:lora_tput}]{
        \includegraphics[width=.35\textwidth]{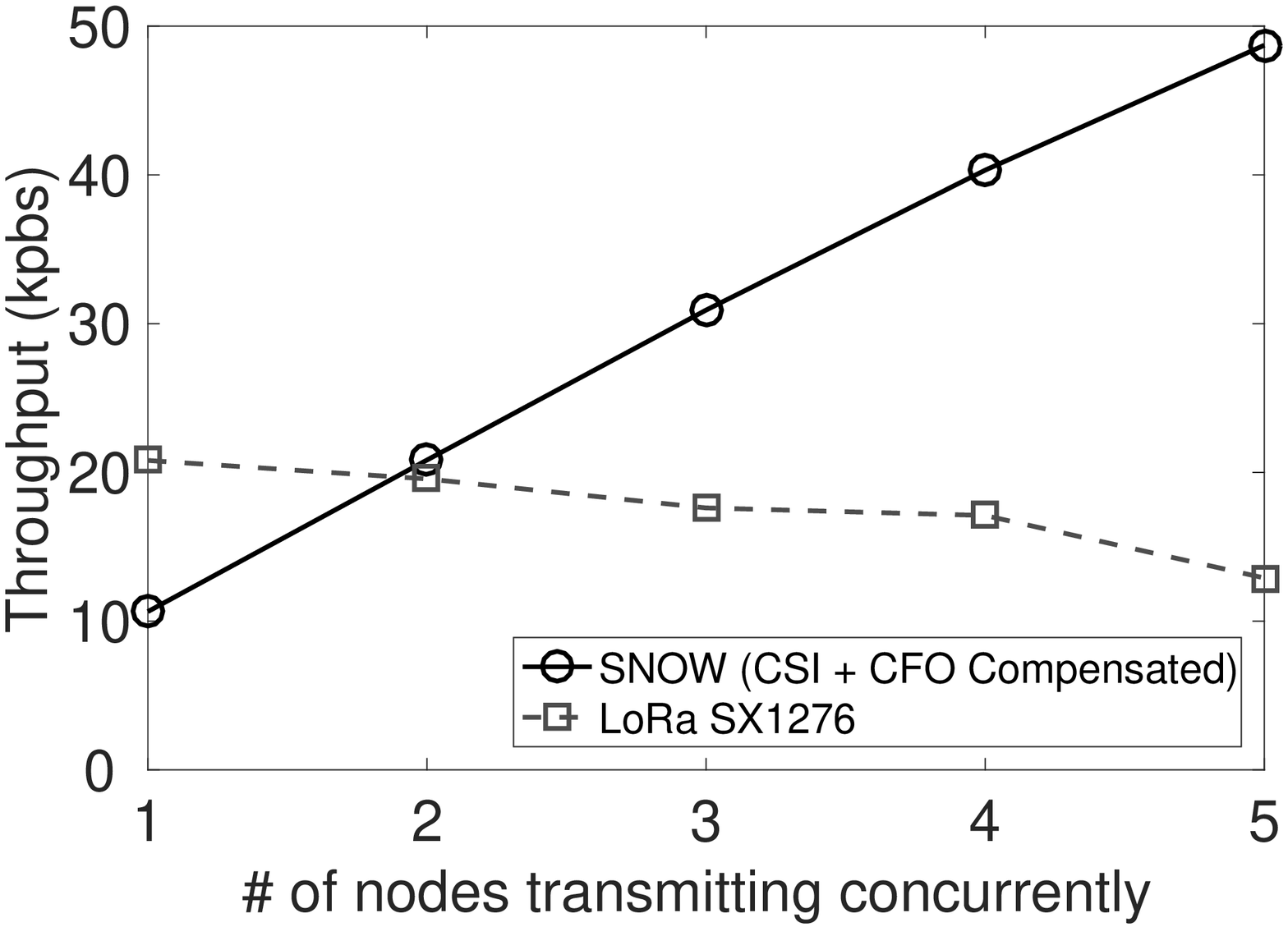}
      }
      \hfill 
      \subfigure[Energy consumption at the nodes\label{fig:lora_en}]{
        \includegraphics[width=.35\textwidth]{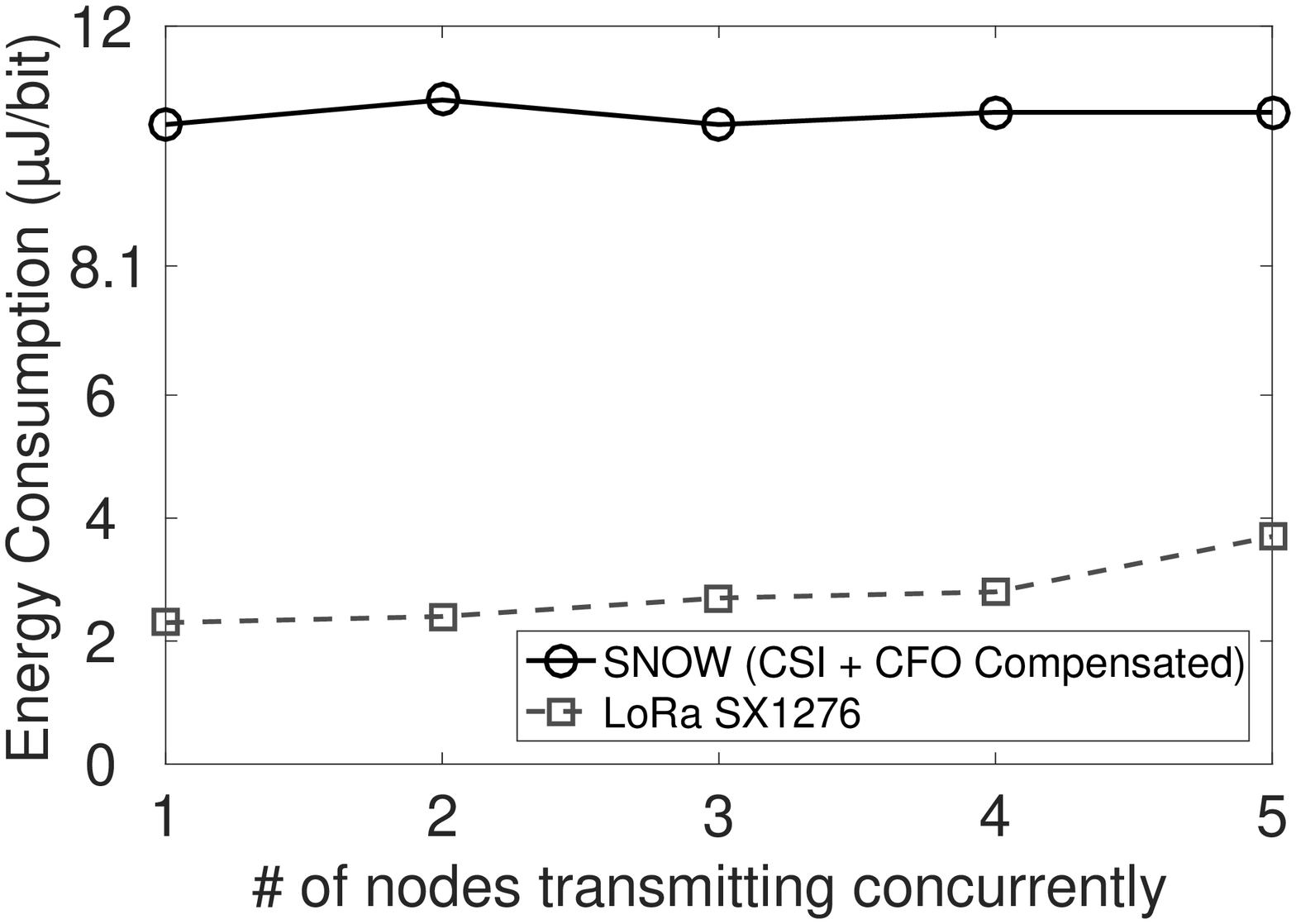}
      }
    \caption{Uplink performance comparison between SNOW and LoRaWAN.}
    \label{fig:lora-comp}
 \end{figure*}
\subsubsection{Reliability Comparison with Parallel Tx/Rx}
Figure~\ref{fig:lora_prr} shows the PRR at the gateway/BS for LoRaWAN and CC13x0-based SNOW implementation under varying number of nodes that transmit concurrently. As shown in this figure, when only one node transmits, the PRR is approximately 95\% in both LoRaWAN and SNOW. Also, the PRR of LoRaWAN decreases with the increase in the number of parallel transmissions. For SNOW, it remains almost similar with the increase in the number of parallel transmissions. For example, when 5 nodes transmit in parallel, LoRaWAN achieves a PRR of 59\%, compared to 87\% in SNOW. Such performance degradation in LoRaWAN happens as it uses an ALOHA-based MAC protocol without any collision avoidance. The PRR of LoRaWAN may increase if we increase the inter-packet interval and will remain the same for SNOW even if we decrease the inter-packet interval. 

\subsubsection{Throughput Comparison}
Figure~\ref{fig:lora_tput} shows the overall throughput (kbps) comparison at the gateway/BS between LoRaWAN and SNOW. As shown in this figure, the throughput at the LoRaWAN gateway is approximately 20.8kbps, compared to 10.64kbps at the SNOW BS when only one node transmits. However, the throughput at the SNOW BS surpasses that at the LoRaWAN gateway when 2 or more nodes transmit concurrently. As shown in Figure~\ref{fig:lora_tput}, the throughput at the SNOW BS is $\frac{48.12}{12.9} \approx 3.7$x higher compared to LoRaWAN when 5 nodes transmit concurrently. The throughput in LoRaWAN decreases as we increase the number of LoRaWAN nodes because of the following reason. The LoRaWAN nodes adopt the ALOHA-based MAC protocol. Thus, a LoRaWAN node does not check if a channel is free before transmitting a packet, which results in a collision with another ongoing packet transmission (if any) from a different LoRaWAN node in the same channel. Consequently, both packets are lost as well as not considered in the throughput calculation.
Packet collisions might happen in our setup since 5 LoRaWAN nodes share 3 channels. Note that our setup is realistic, which emulates the scenario of having hundreds of LoRaWAN nodes under a 64-channel (maximum possible) LoRaWAN gateway. For 100 LoRaWAN nodes, the channel to node ratio is $\frac{64}{100} \approx \frac{3}{5}$, which is also the same in our setup. The SNOW nodes, on the other hand, avoid packet collisions on the same channel by adopting the CSMA/CA MAC protocol.
Compared to LoRaWAN, our CC13x0-based SNOW implementation thus observes better throughput.

\subsubsection{Energy Consumption Comparison}
Figure~\ref{fig:lora_en} shows the per packet energy consumption
at the nodes of LoRaWAN and our CC13x0-based SNOW implementation. As the figure shows, when 5 nodes transmit in parallel, a LoRaWAN node spends 3.7$\mu$J/bit, compared to 10.6$\mu$J/bit in SNOW. Here, the per-bit energy consumption in SNOW is slightly higher than that in LoRaWAN. However, this figure shows that the per-bit energy consumption increases in LoRaWAN and remains almost steady in SNOW when the number of concurrent nodes increases. SNOW is designed to enable a large number of concurrent transmissions to the BS and such a tendency in energy consumption shows its energy efficiency under that scenario. On the other hand, the number of retransmissions to deliver a packet increases with the increase in the number of nodes in LoRaWAN, thereby increasing the per packet energy consumption. Due to a limited number of devices, we are unable to demonstrate this in real experiment. Note that the LoRaWAN configuration used in this experiment (e.g., channel bandwidth of 500kHz, spreading factor of 7, coding rate of $\frac{4}{5}$) is the most energy-efficient (for a given Tx power) for an individual LoRaWAN node. 
As reported in~\cite{xu2019measurement}, increasing the spreading factor increases energy consumption at the LoRaWAN nodes. Similarly, decreasing the channel bandwidth also increases energy consumption at the LoRaWAN nodes~\cite{xu2019measurement}.

\subsection{Discussion on the Performance of CC1310 and CC1350}
As we experiment with both CC1310 and CC1350 devices as the SNOW nodes (Sections~\ref{sec:prr_dist}--\ref{sec:mobility_per}), we see no noticeable performance difference at the SNOW BS in terms of reliability and throughput, compared to our previous experiments with the CC1310 devices only in~\cite{snow_cots}. Similarly, there is no noticeable performance difference in terms of end-to-end latency and energy consumption between the two platforms. 
The reason is that our SNOW implementation is minimally invasive to the IoT devices having almost similar PHY layer properties. We envision that any IoT device with a programmable PHY, capable of operating in the white spaces, and capable of OOK/BPSK modulation may work as a SNOW node by addressing the same set of challenges as described through Sections~\ref{sec:csi}--\ref{sec:near-far}. In this paper, we practically demonstrate this by implementing SNOW on CC1310 and CC1350 devices and showing that their performance is similar.

\section{Related Work}\label{sec:related}
Recently, a number of LPWAN technologies have been developed that operate in the licensed (e.g., LTE Cat M1, NB-IoT, EC-GSM-IoT, and 5G) or unlicensed (e.g., LoRa, SigFox, RPMA (INGENU), IQRF, Telensa,  DASH7, WEIGHTLESS-N, WEIGHTLESS-P, IEEE 802.11ah, IEEE 802.15.4k, and IEEE 802.15.4g) spectrum~\cite{ismail2018low, whitespaceSurvey, saxena2016achievable, chen2017narrowband, gozalvez2016new, akpakwu2017survey, kouvelas2020p}.
Operating in the licensed band is costly due to high service fee and costly infrastructure. 
On the contrary, most non-cellular LPWANs including LoRa and SigFox operate in the sub-1GHz ISM band, for example, between 902 and 928MHz in the continents of North America and South America. While the ISM band is unlicensed, it is heavily crowded due to the proliferation of LPWANs as well as other wireless technologies in this band. 
To avoid the high cost of licensed band and the crowd of the ISM band, SNOW was designed to exploit the TV white spaces. As reported in~\cite{whitespaceSurvey, ismail2018low, snow2, ws_sigcomm09},
white spaces are widely available in both urban and rural areas, less crowded (compared to the ISM band), and offer a wider spectrum (compared to the available frequency bands for LPWANs in the ISM band) that a SNOW BS can utilize.

The existing work on white space focused on exploiting the white spaces for broadband access~\cite{whitespaceSurvey, zhang2015design, hasan2014gsm, kumar, harrison2015whitespace, ws_sigcomm09, WATCH, videostreaming, linkasymmetry, vehiclebased, ws_mobicom13, ws_nsdi10} and spectrum determination through spectrum sensing~\cite{saeed2017local, ws_dyspan08_kim, ws_mobicom08, ws_dyspan11, FIWEX} and/or geo-location database approach~\cite{database1, database2, database3, vehiclebased, hysim}. 
Alongside, various standards bodies (e.g., IEEE 802.11af, IEEE 802.15.4m, IEEE 802.19.1, IEEE 802.22, IEEE 1900.4a, IEEE 1900.7, and ECMA-392) and industry leaders (e.g., Microsoft and Google) have also targeted the white spaces for unlicensed personal or commercial use~\cite{kocks2012spectrum, MSRAfrica, ismail2018low, whitespaceSurvey}. In contrast, SNOW, as a LPWAN, exploits white spaces for highly scalable and wide-area sensor network applications.
With the rapid growth of IoT,  LPWANs will suffer from crowded spectrum due to long range. It is hence critical to exploit white spaces for IoT. 
Our paper focuses on implementing SNOW using the cheap and widely available COTS devices for practical and scalable deployment.

\section{Conclusions}\label{sec:conclusion}
The recently proposed LPWAN technology -- SNOW -- has the potential to enable connectivity to numerous IoT devices over long distances. However, the high cost and the large form-factor of the USRP-based SNOW nodes hinder its practical deployments. In this paper, we have implemented SNOW for practical deployments using the CC13x0 devices as SNOW nodes. Our CC13x0-based SNOW implementation decreases the cost and the form-factor of a single SNOW node by 25x and 10x, respectively. 
We have also addressed several practical deployment challenges that include PAPR reduction, CSI estimation, CFO estimation, and near-far power problem. 
We have deployed our CC13x0-based SNOW in the city of Detroit, Michigan and achieved per node uplink and downlink throughputs of 11.2kbps and 4.8kbps, respectively, over a distance of 1km. 
Our experiments also show that SNOW can achieve throughput several times higher than LoRaWAN under typical settings.
Finally, our extensive experiments have demonstrated the CC13x0-based SNOW as a feasible LPWAN technology that can be deployed practically at low-cost and in large-scale for future IoT applications.
\begin{acks}
This work was supported by NSF through grants CAREER-1846126 and CNS-2006467.
\end{acks}

\bibliographystyle{ACM-Reference-Format}
\bibliography{whitespacebib}

\end{document}